\documentclass[10pt,journal,compsoc]{IEEEtran}
\ifCLASSOPTIONcompsoc
  \usepackage[nocompress]{cite}
\else
  \usepackage{cite}
\fi
\ifCLASSINFOpdf
  \usepackage[pdftex]{graphicx}
\else
\fi
\graphicspath{{figures/}{pictures/}{images/}{./}{fig/}} 

\usepackage{microtype}                 
\usepackage{textcomp}                  
\usepackage{mathptmx}                  
\usepackage{times}                     
\usepackage{cite}                      
\usepackage{tabu}                      
\usepackage{booktabs}                  

\usepackage[table]{xcolor}

\usepackage{color}
\usepackage{transparent}
\usepackage{comment}
\usepackage{amsfonts}
\usepackage{enumitem}
\usepackage[normalem]{ulem}
\usepackage{colortbl}
\usepackage{graphicx}
\usepackage{multirow}
\usepackage{mathrsfs}
\usepackage{mathtools}
\usepackage{amsmath,amsfonts,amssymb,amsthm}
\usepackage{mathtools}
\usepackage{commath}
\usepackage[sc,osf]{mathpazo}
\usepackage[utf8]{inputenc}
\usepackage{tikz}
\usepackage{graphicx}

\usepackage{hyperref}
\usepackage{comment}
\usepackage{multirow}
\usepackage{color}
\usepackage{rotating}
\usepackage{setspace}
\usepackage{xspace}
\usepackage{wrapfig}
\usepackage{subcaption}
\usepackage{adjustbox}
\hypersetup{hidelinks}

\definecolor{red}{rgb}{0.6, 0, 0}
\definecolor{blue}{rgb}{0, 0, 0.6}
\definecolor{green}{rgb}{0.16, 0.435, 0.16}

\usepackage[utf8]{inputenc}

 \newcommand{\textremoved}[1]{\textcolor{red}{\sout{#1}}}
 \newcommand{\removeditem}[1]{\item \textremoved{#1}}
 \newcommand{\camerareadyonly}[1]{}

 \renewcommand{\textremoved}[1]{}
\renewcommand{\removeditem}[1]{}
 \renewcommand{\camerareadyonly}[1]{#1}

  \newcommand{\mycbox}[1]{\tikz{\path[draw=#1,fill=#1] (0,0) rectangle (0.3cm,0.3cm);}}

\newcommand{\para}[1]{\textbf{#1.}\xspace}

\definecolor{red}{rgb}{0.6, 0, 0}
\definecolor{blue}{rgb}{0, 0, 0.6}
\definecolor{green}{rgb}{0.16, 0.435, 0.16}

\makeatletter
\let\orgdescriptionlabel\descriptionlabel
\renewcommand*{\descriptionlabel}[1]{%
  \let\orglabel\label
  \let\label\@gobble
  \phantomsection
  \edef\@currentlabel{#1\unskip}%
  \let\label\orglabel
  \orgdescriptionlabel{#1}%
}
\makeatother

\newcommand{\doubleBlindText}[1]{\textit{$<$Withheld for double-blind review$>$}}

\usepackage{graphicx}

\begin{document}

\title{Automatic Scatterplot Design Optimization for Clustering Identification}
%



\author{Ghulam Jilani Quadri, Jennifer Adorno Nieves, Brenton M. Wiernik and Paul Rosen
\IEEEcompsocitemizethanks{\IEEEcompsocthanksitem G.J.\ Quadri was with the Department of Computer Science, University of North Carolina, Chapel Hill,
NC, 27599.\protect\\
E-mail: ghulamjilani@usf.edu

\IEEEcompsocthanksitem J.A.\ Nieves was with the Department
of Computer Science and Engineering, University of South Florida, Tampa,
FL, 33620.\protect\\
E-mail: jorgea1@usf.edu

\IEEEcompsocthanksitem B. M. Wiernik was with the Department
of Psychology, University of South Florida, Tampa,
FL, 33620.\protect\\
E-mail: brenton@wiernik.org

\IEEEcompsocthanksitem P.\ Rosen was with the Scientific Computing and Imaging Institute, University of Utah, Salt Lake City,
UT, 84112.\protect\\
E-mail: prosen@sci.utah.edu}
\thanks{Manuscript received April 19, 2005; revised August 26, 2015.}}

%
%

\markboth{Journal of \LaTeX\ Class Files,~Vol.~14, No.~8, August~2015}%
{Shell \MakeLowercase{\textit{et al.}}: Bare Demo of IEEEtran.cls for Computer Society Journals}
%




\IEEEtitleabstractindextext{%
\begin{abstract} 
    Scatterplots are among the most widely used visualization techniques. Compelling scatterplot visualizations improve understanding of data by leveraging visual perception to boost awareness when performing specific visual analytic tasks. Design choices in scatterplots, such as graphical encodings or data aspects, can directly impact decision-making quality for low-level tasks like clustering. Hence, constructing frameworks that consider both the perceptions of the visual encodings and the task being performed enables optimizing visualizations to maximize efficacy. In this paper, we propose an automatic tool to optimize the design factors of scatterplots to reveal the most salient cluster structure. Our approach leverages the merge tree data structure to identify the clusters and optimize the choice of subsampling algorithm, sampling rate, marker size, and marker opacity used to generate a scatterplot image. We validate our approach with user and case studies that show it efficiently provides high-quality scatterplot designs from a large parameter space.
\end{abstract}

}

\maketitle
%






\setstretch{1}

\IEEEraisesectionheading{\section{Introduction}\label{sec:introduction}}

\IEEEPARstart{S}{catterplots} are an intuitive and widely used visualization for bivariate quantitative data that can reveal relationships and patterns between the variables~\cite{friendly2005early}. Several studies have evaluated the effectiveness of scatterplots in low-level tasks\cite{quadri2021survey}, that include assessing trends~\cite{nguyen2016visual}, correlation perception~\cite{rensink2010perception, harrison2014ranking}, average values and relative mean judgments~\cite{gleicher2013perception}, detecting outliers~\cite{sarikaya2018design}, and clustering~\cite{quadri2020modeling, munzner2014visualization}.

Design choices in scatterplots, including both the visual encodings, e.g., data point size or opacity, and data aspects, e.g., the number of data points, directly impact the quality of decision-making for low-level tasks~\cite{amar2005low}. Effective visualization design enhances comprehension by leveraging visual perception. Several studies have focused on optimizing a scatterplot by adjusting data point size~\cite{kim2018assessing}, the number of data points~\cite{gleicher2013perception}, opacity~\cite{micallef2017towards}, color~\cite{szafir2018modeling}, and shape~\cite{sedlmair2012taxonomy}.

One particular problem for scatterplots is overplotting, which occurs when many data points overlap and obscure the underlying data patterns. A combination of design choices can be made to reduce its influence, including choosing a subsampling algorithm and adjusting the sampling rate, reducing mark size or opacity, or some combination of both~\cite{ellis2007taxonomy}. A designer's control over design elements that influence overplotting varies from complete control, e.g., point size and opacity, to limited control, e.g., the number of data points via subsampling, to no control, e.g., the distribution of points, which are inherent to the data. Given the number of factors designers control, the design space is large for a manual search, as an optimal design must consider the influence each parameter has on the others and the task being performed when recommending design choices.

\begin{figure*}[!t]
    \centering
    \includegraphics[width=0.995\linewidth]{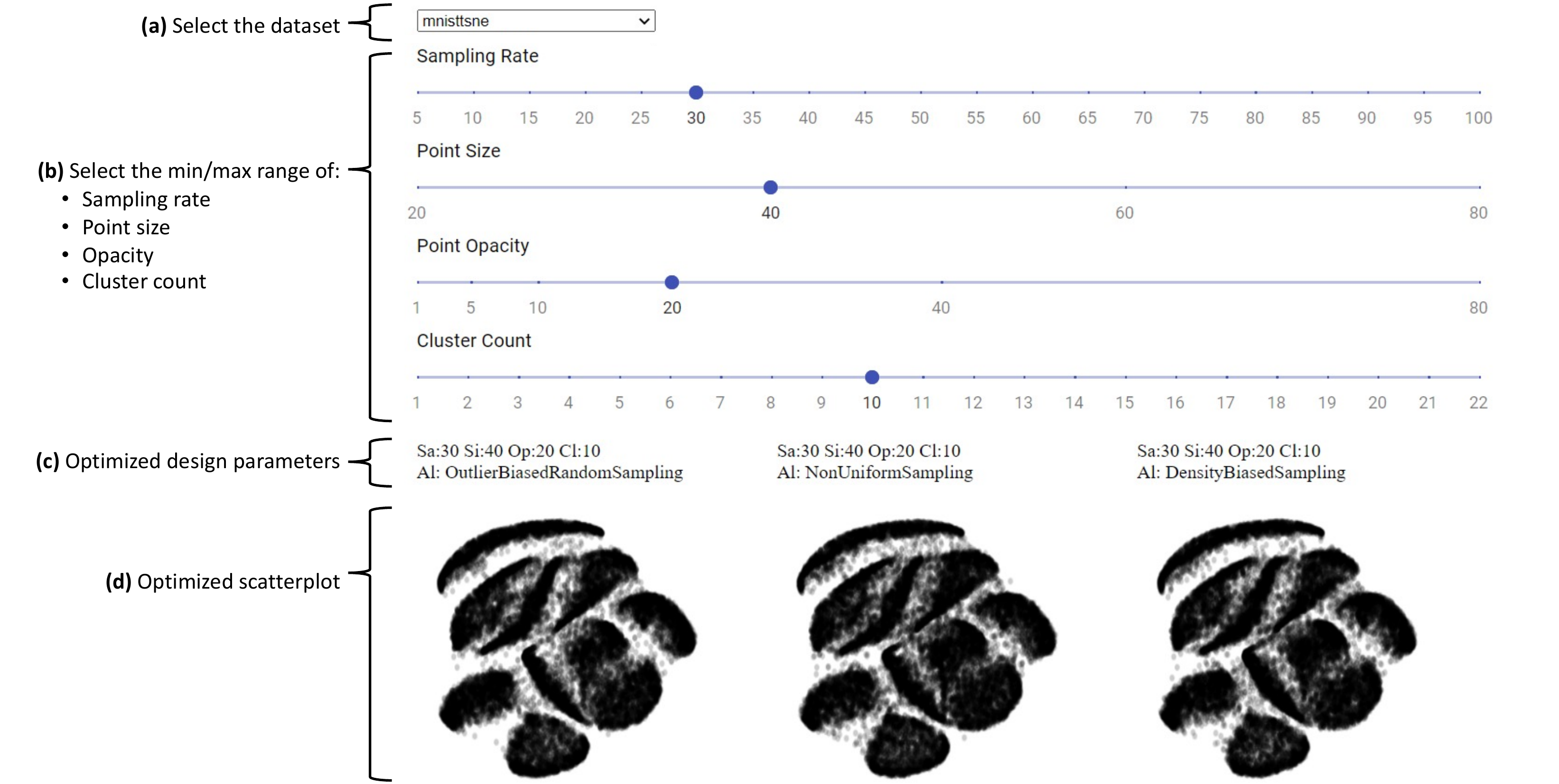}
    \caption{Example of identifying an optimal design using our approach. (a)~The user first selects a dataset. (b)~Optionally, the user selects a minimum and maximum range of the parameters (in this example, the sliders are set to the same minimum/maximum values; see our demo for examples of setting ranges), including sampling rate (\% of data), point size (area of circle in pixels), point opacity (\% alpha), and cluster count, which the user can limit but is not optimized by our approach. (c)~Our approach presents scatterplot parameters that optimize cluster saliency, ordered by saliency values 0.043847, 0.039364, and 0.021686, from left to right. (d)~The scatterplots associated with the design parameters are shown.} 
    \label{fig:interactivemodel}
\end{figure*}

In this paper, we consider the problem of automatic design optimization in scatterplots for the task of \textit{clustering}. Clustering occurs when patterns in the data form distinct groups~\cite{amar2005low,sarikaya2018design}. However, while identifying clustering structure is generally considered an ill-defined problem, Quadri and Rosen\cite{quadri2020modeling} recently introduced a model for accurately capturing human perception of different numbers of clusters in scatterplots using methods from Topological Data Analysis~\cite{quadri2020modeling}. The method encodes the information into a \textit{threshold plot}, calculated on the visual density to measure the visibility of different numbers of clusters in a scatterplot.

In this paper, we extend their work by utilizing the threshold plot for scatterplot design optimization. We first define a saliency measure on the threshold plot to rank the scatterplots by how salient their cluster structure is. We then evaluate an input scatterplot on the parameters that influence visual density, including data aspects, i.e., subsampling algorithm and sampling rate, and visual encodings, i.e., mark size and opacity. Finally, our approach automatically optimizes visualization designs by ranking them from highest to lowest in terms of cluster task performance.

Our approach is implemented into an open-source web tool (see \autoref{fig:interactivemodel}). We validated it through a user study conducted on 70 participants from Amazon Mechanical Turk (AMT). We found that the saliency of the threshold plot is a good proxy for cluster structure when selecting an optimal scatterplot design. The effect was particularly pronounced when the value for the saliency was high. Further, a case study showed that our approach requires less interaction and time to select an optimal design over a manual search.

\textbf{Contributions:} The contributions of work are (1)~an \textit{optimization model} for parameters that ranks combinations of parameters using a saliency measure for the task of cluster analysis; (2)~an open-source web tool that can be deployed to propose optimal designs for an input scatterplot based on its clustering structure; and (3)~an evaluation of the approach with a user study including 70 subjects and a case study involving 10 visualization students.

\section{Prior Work}

    We provide brief coverage of clustering, design optimization, and overdraw reduction.

\subsection{Clustering in Scatterplots}

    Clustering, broadly defined, is the ``grouping of similar data points on scatterplot in a given dataset''~\cite{amar2005low} to reveal characteristics of data and allow further exploration of underlying patterns~\cite{sarikaya2018scatterplots, sarikaya2018design}. Previous works have investigated modeling cluster perception in scatterplots. Aupetit et al.\ studied how 1,400 variants of clustering algorithms matched human impressions of clustering structure in scatterplots and found CLIQUE, DBSCAN, and Agglomerative clustering each captured some aspects of human perception~\cite{aupetit2019toward}. Matute et al.'s technique quantified and represented scatterplots through skeleton-based descriptors measuring scatterplot similarity~\cite{matute2017skeleton}. However, their approach does not consider visual encodings in the evaluation. Sedlmair and Aupetit developed an approach to mimic human judgment of class separation by using machine learning on 15 class separation measures on scatterplots~\cite{sedlmair2015data}.
    ScatterNet, a deep learning model, captures perceptual similarities between scatterplots that could be used to emulate human clustering decisions~\cite{ma2018scatternet}. Scagnostics focused on identifying patterns in scatterplots, including clusters~\cite{dang2014transforming}, but Pandey et al.\ later showed they do not reliably reproduce human judgments~\cite{pandey2016towards}.

\subsection{Design Optimization in Scatterplots}

    Rensink's framework for reasoning about perception of visualization designs suggests using techniques from vision science~\cite{rensink2014prospects}. The extended-vision theory asserts that a viewer and visualization are a single system, whereas the optimal-reduction thesis postulates the existence of an optimal visualization. The work focuses on the fundamental question of \textit{can we determine if a design is optimal?}

    Optimization studies have focused on several aspects of scatterplots, including color assignment in scatterplot design to optimize class separability taking into account density-related factors, such as spatial relationship, density, degree of overlaps between points and cluster, and background-color~\cite{wang2018optimizing}; creating specialized color pallets that help with visual separation of classes in multi-class data~\cite{lu2020palettailor}; automatically selecting the optimal representation between scatterplot and line graph for trend exploration in time series data~\cite{wang2018line}; and perceptual optimization of scatterplot design on standard design parameters, including mark size, opacity, and aspect ratio, demonstrating effective choices of those variables to enhance class separation~\cite{micallef2017towards}.
   
    Recently, ClustMe used visual quality measures (VQMs), which algorithmically emulate human judgments, to model human perception to rank scatterplots~\cite{abbas2019clustme}. It performed well in reproducing human decisions for cluster patterns. Their perceptual data was later used to build a model evaluating how far existing techniques of VQM align with clusters perceived by humans~\cite{aupetit2019toward}.  In another study, 15 state-of-the-art class separation measures were evaluated, and human ground truth data on color-coded 2D scatterplots was used to learn how well a measure would predict human judgments on previously unseen data~\cite{sedlmair2015data}. 
    
    In the prior work of Quadri and Rosen, a \textit{threshold plot} was generated using visual density of a scatterplot to model the number of clusters visible for a given set of design factors~\cite{quadri2020modeling}. In this work, we extend the application of threshold plots to instead use them for selecting the optimal design to enhance user perception of the cluster structure in a scatterplot.

\subsection{Overdrawing in Scatterplots and Solutions}   
    
    Overplotting, the over-saturation of visual density, in scatterplots makes data analysis inefficient by obscuring underlying data patterns. A taxonomy of clutter-reduction techniques~\cite{ellis2007taxonomy} suggests several approaches for reducing clutter, including varying mark size~\cite{bederson2002ordered, derthick2003constant, plaisant1996lifelines, woodruff1998constant}, varying opacity level~\cite{matejka2015dynamic,wegman1997high, kosara2002focus+, johansson2006revealing, fekete2002interactive}, and subsampling data points~\cite{cohen1973eta,chen2014visual,urribarri2017prediction,hu2019data}.

\subsubsection{Reducing Point Size}

    The size of marks in a scatterplot is an important factor in visual aggregation tasks~\cite{szafir2016four}. As the size of data points on the scatterplot increases, so does the density, which directly influences cluster perception~\cite{sadahiro1997cluster}. Scatterplot designs with larger points may obscure the visibility of underneath points, and hence reducing the point size would be beneficial. However, reducing the point size can conflict with color-based encodings on the data points as color difference varies with point size~\cite{szafir2018modeling}. We consider monochrome scatterplots in our case. Therefore, we do not observe such a conflict. In the case of relatively minor overplotting, reducing point size can be helpful, but when point size is already as small as possible, i.e., 1 pixel, this method cannot be used~\cite{few2008solutions}.

\subsubsection{Reducing Point Opacity}
    Reducing mark opacity can alleviate overplotting to assist visual analytics tasks~\cite{sarikaya2018scatterplots, quadri2020modeling}, e.g., spike detection in dot plots~\cite{correll2018looks}. Furthermore, varying opacity levels aid in different visual tasks---while low opacity benefits density estimation for large data, it also makes locating outliers more challenging~\cite{micallef2017towards}. Matejka et al.\ defined an opacity scaling model for scatterplots based on the data distribution and crowdsourced responses to opacity scaling tasks~\cite{matejka2015dynamic}. Although a change in opacity cannot avoid overlap, it can reveal a small number of underlying or partially overlapping points or overview behavior of points~\cite{ellis2007taxonomy}. Further, making the points more transparent is less helpful when there are many points.

\subsubsection{Data Subsampling} 
\label{sec-label:relatedsampling}
   
    \para{Sampling Rate}The quantity of data points on the screen directly influences the visual density and overdrawing of a scatterplot. Gleicher et al.'s empirical study asked participants to compare and identify average values in multi-class scatterplots~\cite{gleicher2013perception}. It demonstrated that judgments are improved with a higher number of points. Also, the number of data points affects the user's performance on cluster perception in a given scatterplot~\cite{quadri2020modeling}. Reducing the number of points reduces the overplotting and reveals underlying patterns~\cite{ellis2007taxonomy}. 

    \para{Sampling Algorithm} The simplest way to reduce the number of points is to randomly sample the data, which preserves dense cluster regions but may lose low-density ones~\cite{ellis2002density, dix2002chance}. Bertini and Santucci modeled the relationship between the visual density and clutter, which could be used to determine the right sampling ratio, and presented an automatic method to preserve the relative densities~\cite{bertini2005improving}. Improvements to the random method use non-uniform sampling that treats parts of the scatterplot differently to preserve certain properties~\cite{yuan2020evaluation}. In \autoref{sec-label:sampling}, we discuss several techniques that preserve relative visual density between clusters, preserve outliers when subsampling, or preserve the spatial separation between clusters.

\subsubsection{Density-based Data Representations}

    There have been several variations on scatterplots that utilize alternative density representations to overcome overplotting. Carr et al.\ used hexagonal cells to accumulate densities~\cite{carr1987scatterplot}. Bachthaler and Weiskopf created a continuous density field using a mathematical model to produce the continuous scatterplots~\cite{bachthaler2008continuous}. Keim et al.\ developed the generalized scatterplot, which allows users to balance overplotting and distortion~\cite{keim2010generalized}. Mayorga and Gleicher proposed Splatterplots, which showed dense regions as smooth contours and  discrete markers to highlight outliers~\cite{mayorga2013splatterplots}. A recent study, called Sunspot Plots, demonstrated that a smooth blending of discrete and continuous representations enables the visualization of leading trends in dense areas while still preserving outliers in sparse regions~\cite{trautner2020sunspot}.

\section{Methods}
\label{sec-label:methodology}

\begin{figure}[!b]
    \centering
    
    \includegraphics[width=0.995\linewidth]{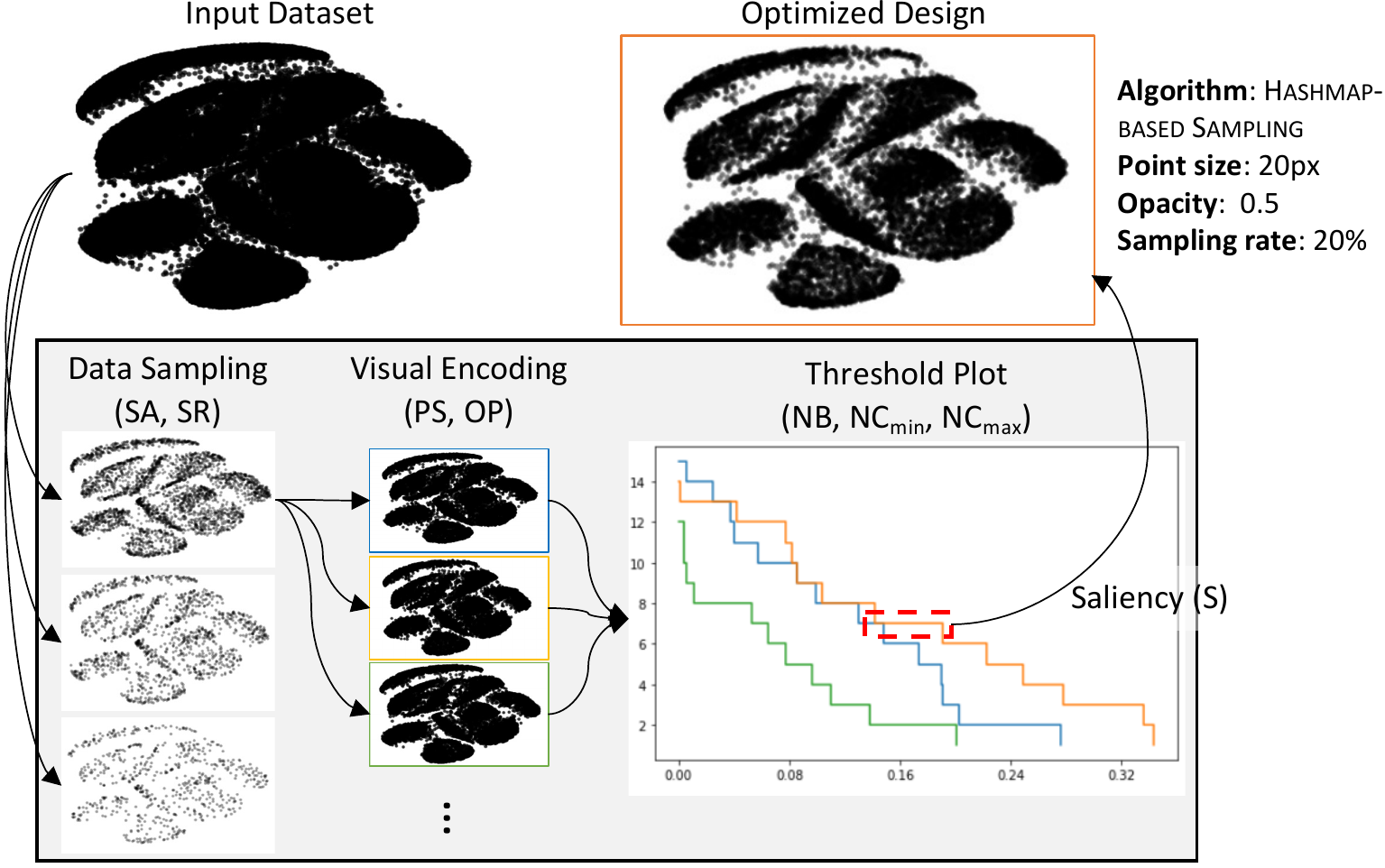}
    \caption{Illustration of our approach. The input data (MNIST) are processed through four stages: data sampling (\autoref{sec-label:sampling}), visual encoding (\autoref{sec-label:encoding}), calculation of a threshold plot (\autoref{sec-label:threshold}), and finally, an optimized design is presented by ordering the saliency measure (in the red box) of scatterplots from highest to lowest (\autoref{sec:optimize}).}
    \label{fig:overview}
\end{figure}

    Visualization effectiveness is a task-dependent engagement directly impacted by the design choices. Our objective is to provide design choices for a scatterplot when optimizing for cluster structure saliency. Our approach allows interactively choosing the optimal design through a user-guided automatic parameterization that uses a \textit{threshold plot}~\cite{quadri2020modeling} to model cluster perception. The optimized parameterization of the scatterplot considers data aspects, including the number of data points, and visual encodings, including data point size and opacity. The output is a set of scatterplots ranked by their \textit{cluster saliency} from highest to lowest.

    As an overview of the process, the data are input into the following processing stages, as shown in \autoref{fig:overview}.

\vspace{5pt}
\begin{wrapfigure}[2]{L}{0.05\linewidth}         
\vspace{-11pt}
    \begin{minipage}[t]{1.2\linewidth}
        \includegraphics[width=20pt]{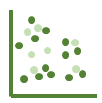}
    \end{minipage}
\end{wrapfigure}
\noindent
\uline{Sampling} (\autoref{sec-label:sampling}): Data are first subsampled with different numbers of points or sampling rate using various algorithms.

\vspace{2pt}
\begin{wrapfigure}[2]{L}{0.05\linewidth}         \vspace{-11pt}
    \begin{minipage}[t]{1.2\linewidth}
        \includegraphics[width=20pt]{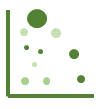}
    \end{minipage}
\end{wrapfigure}
\noindent
\uline{Visual Encoding} (\autoref{sec-label:encoding}): A variety of data point size and opacity values are used to encode the points.

\vspace{2pt}
\begin{wrapfigure}[2]{L}{0.05\linewidth}            \vspace{-11pt}
    \begin{minipage}[t]{1.2\linewidth}
        \includegraphics[width=20pt]{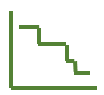}
    \end{minipage}
\end{wrapfigure}
\noindent
\uline{Threshold Plot} (\autoref{sec-label:threshold}): The visual density of the scatterplot is calculated and \textit{threshold plots} are constructed.
    
\vspace{2pt}
\begin{wrapfigure}[2]{L}{0.05\linewidth}            \vspace{-11pt}
    \begin{minipage}[t]{1.2\linewidth}
        \includegraphics[width=20pt]{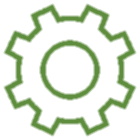}
    \end{minipage}
\end{wrapfigure}
\noindent
\uline{Optimized Design} (\autoref{sec:optimize}): Finally, a saliency measure is extracted from the threshold plots, which are then ranked from highest to lowest saliency and presented to the user as the optimized design choice.

\begin{figure*}[!t]
        \centering
        \includegraphics[width=0.975\linewidth]{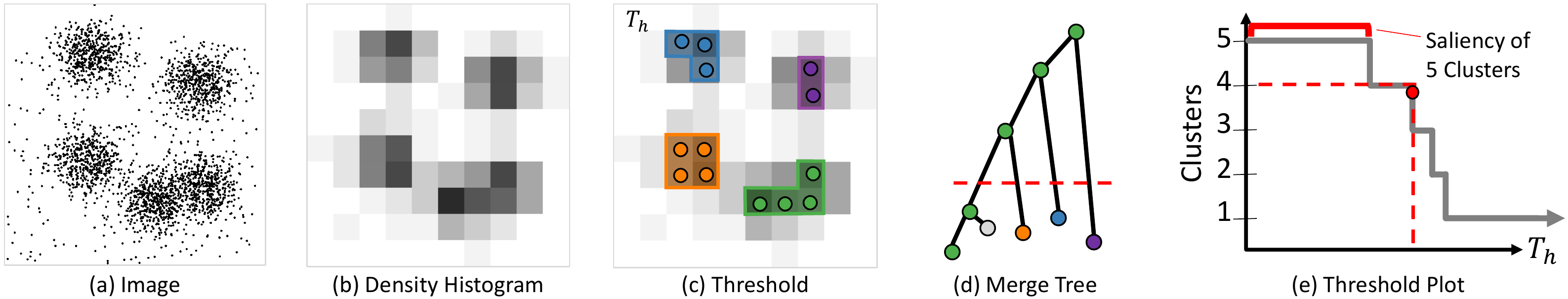}
     
        \caption{Illustration of generating a \textit{threshold plot} (refer to \cite{quadri2020modeling} for details) to computes a \textbf{saliency score}. (a)~A scatterplot is given as input. (b)~A density histogram (of $20 \times 20$ in our implementation) is calculated. (c)~The histogram is evaluated at different density thresholds, and components are extracted. (d)~A merge tree is created by tracking components across thresholds. (e)~A threshold plot is generated from the merge tree. The horizontal axis represents the a persistence threshold of clusters, while the vertical axis shows the number of clusters visible at that threshold. The dashed red line shows how a threshold can be extracted from a given number of clusters and vice versa. Finally, the saliency for some number of clusters is measured as the range of the minimum and maximum threshold for that number of clusters.}
        \label{fig:Thresholdplot_generation}
\end{figure*}

\subsection{Data Sampling}
\label{sec-label:sampling}

\begin{wrapfigure}[3]{L}{0.09\linewidth}         \vspace{-11pt}
\begin{minipage}[t]{1.2\linewidth}
        \includegraphics[width=33pt]{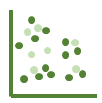}
        \end{minipage}
\end{wrapfigure} 
Data subsampling is dependent on sampling algorithm (SA) and sampling rate (SR). As discussed in \autoref{sec-label:relatedsampling}, a good quality subsampling algorithm decreases the visual clutter by reducing the number of data points while retaining some of the original structure of the data. However, the best algorithms often turn out to be time-intensive to compute. Our approach considers a collection of many algorithms at a variety of sampling rates to identify an optimal one. We organize the subsampling techniques in \autoref{tbl:sampling} based on the properties they preserve, including \textit{random}, \textit{relative visual density preserving}, \textit{outlier preserving}, and \textit{spatial separation preserving}.  Though all sampling rates are used by default, users may optionally select a subset of sampling rates to use.

\subsubsection{Random sampling} 

\textsc{Random} sampling is a classic method for revealing structures in data~\cite{rojas2017sampling}. \textsc{Random} sampling works by selecting output samples with equal probability. Example studies using \textsc{Random} sampling include those by Ellis and Dix~\cite{ellis2002density, dix2002chance}.

\uline{Advantages \& Limitations:} \textsc{Random} sampling does not require special knowledge of the data and is widely available in existing tools. It preserves relative intensity differences, but since points are removed with equal probability, cluster structure may disappear, sampling artifacts can be introduced, and outliers may or may not be preserved.

\subsubsection{Preserving Relative Visual Density}

The next type of sampling methods aim to preserve the visual density of both dense and sparser regions. \textit{Visual (data) density} is the ratio between the number of displayed data samples and their corresponding rendered area.  \textit{Density preserving} algorithms optimize the weights of each sampled group to be proportional to the original group's size~\cite{palmer2000density}.

\textsc{Density Biased} sampling works by probabilistically over-sampling sparse regions and under-sampling dense regions~\cite{palmer2000density}, thus preserving small clusters and solitary samples while reducing sampling in dense regions. \textsc{Non-uniform} sampling strategies assign varying sampling probability to data so that some specific properties of the data can be better preserved~\cite{bertini2006give, bertini2004chance}. The approaches divide the sample space into a uniform grid, determines the density of each grid cell, and selects samples from cells according to their density.  \textsc{SVD}-based sampling formulates visual density preserving as a matrix decomposition solved with singular value decomposition (SVD)~\cite{joia2015uncovering}. This method performs SVD on the original dataset and selects the samples with the most significant correlation with top-$k$ basis vectors. \textsc{Multi-view Z-order} sampling is a density preserving method, formulated as a set cover problem by segmenting Z-order curves of the samples in each class and the whole dataset~\cite{zheng2013quality}. This strategy greedily selects samples that minimize kernel density estimation error~\cite{hu2019data}. \textsc{Recursive Subdivision} sampling is a multi-class scatterplot sampling strategy to preserve relative densities, maintain outliers, and minimize visual artifacts~\cite{chen2019recursive}. It splits the visual space with a KD-tree and determines which class of instances should be selected at each leaf node based on a backtracking procedure.

\begin{table}[!t]
    \centering
    \caption{Sampling algorithm used in our study. Category color-coding denotes the types of features preserved. }
    \label{tbl:sampling}
    \begin{adjustbox}{width=0.995\linewidth}
    \begin{tabular}{| l | c |c@{ }c@{ }c@{ }c|}
        \hline
        Sampling Methods & Application & \multicolumn{4}{c|}{Category} \\
        \hline
        \hline

        \hspace{2pt}  \textsc{Random}                    & \cite{xia2017ldsscanner,dos2012ilamp, poco2011framework,rieck2015persistent} &  \mycbox{gray} & & & \\ 
        \hspace{2pt} \textsc{Density Biased}             & \cite{xiang2019interactive, palmer2000density}  & & \mycbox{red} & & \\ 
        \hspace{2pt}  \textsc{Non-uniform}               & \cite{bertini2004chance,bertini2006give} &   & \mycbox{red} & &  \\ 
        \hspace{2pt}  \textsc{SVD}                 & \cite{joia2015uncovering} &    & \mycbox{red} & &    \\ 
        \hspace{2pt}  \textsc{Multi-view Z-order}        & \cite{hu2019data}  &   & \mycbox{red} & &  \\ 
        \hspace{2pt}  \textsc{Recursive Subdivision}     & \cite{chen2019recursive} & & \mycbox{red} & \mycbox{blue} &  \\
        \hspace{2pt}  \textsc{Outlier Biased Density}    & \cite{xiang2019interactive} & & \mycbox{red} & \mycbox{blue}  & \\ 
        \hspace{2pt}  \textsc{Outlier Biased Random}     & \cite{liu2017visual,zhao2019oui} & & & \mycbox{blue}  &  \\ 
        \hspace{2pt}  \textsc{Hashmap}             & \cite{cheng2018colormap} & & & \mycbox{blue}  &  \\ 
        \hspace{2pt}  \textsc{Outlier Biased Blue Noise} & \cite{xiang2019interactive} & & & \mycbox{blue} & \mycbox{green} \\
        \hspace{2pt}  \textsc{Blue Noise}                & \cite{wei2008parallel,chen2014visual,xiang2019interactive}  & & & & \mycbox{green}\\ 
        \hspace{2pt}  \textsc{Multi-class Blue Noise}    & \cite{chen2014visual,wei2010multi} & & & & \mycbox{green}\\ 
        \hspace{2pt}  \textsc{Farthest Point}            & \cite{berger2016cite2vec}  & & & & \mycbox{green} \\ 
        \hspace{2pt}  \textsc{Z-order}                   & \cite{zheng2013quality,hu2019data} & & & & \mycbox{green} \\ 
        \hline
    \end{tabular}
    \end{adjustbox}    

    \vspace{2pt}
    \begin{adjustbox}{width=0.975\linewidth}
    \footnotesize
    \begin{tabular}{ll}
        \mycbox{gray} Random              & \mycbox{red}  Preserving relative visual density \\
        \mycbox{blue} Preserving outliers & \mycbox{green} Preserving spatial separation     \\
    \end{tabular}
    \end{adjustbox}    
    
    
\end{table}

\uline{Advantages \& Limitations:} These approaches reduce density in overdrawn regions while minimizing decreases in sparse areas. The notable feature of this category is maintaining and preserving the relativeness in the visual density of both dense and sparse regions. However, it can result in substantial cluster pattern disappearance, i.e., reduced cluster separation, and some of the algorithms are time-intensive in terms of computations, e.g., with \textsc{Multi-view Z-order} (see \autoref{fig:timeanalysis}).

\subsubsection{Preserving Outliers}

Preserving outliers is another general goal in sampling strategies. Having no clear definitions, data points in low-density areas are often regarded as outliers~\cite{breunig2000lof}. A typical method for achieving this goal is to update existing sampling algorithms to make them accept more outliers~\cite{liu2017visual, xiang2019interactive}. 

\textsc{Outlier Biased Random} sampling assigns higher sampling probabilities to outliers in random sampling~\cite{liu2017visual}. Other sampling methods have also been adapted to bias their sampling towards outliers, e.g., \textsc{Outlier Biased Blue Noise} sampling~\cite{liu2017visual} and \textsc{Outlier Biased Density} sampling~\cite{xiang2019interactive}. \textsc{Hashmap}-based stratified sampling technique preserves outliers while keeping the main distribution by sampling the point clouds on display using a color mapping~\cite{cheng2018colormap}.

\uline{Advantages \& Limitations:} Preserving outliers conflicts with many of the goals of emphasizing cluster separation. For example, preserving outliers will distort the relative data densities since relatively more data points are selected in low-density regions instead of high-density ones, which will increase the ambiguity between cluster boundaries.

\subsubsection{Preserving Spatial Separation} 

There are some cases where spatial separation between classes/clusters or highly dense regions is desirable.

\textsc{Blue Noise} sampling, inspired by~\cite{yellott1983spectral}, randomly selects samples but remains spatially uniform~\cite{yan2015survey, wei2008parallel}. \textsc{Multi-class Blue Noise} is a multi-class extension that maintains the blue noise properties of each class and of the whole dataset~\cite{wei2010multi}. \textsc{Farthest Point} sampling selects samples with better spatial separation by randomly selecting the initial sample and iteratively selecting additional samples of maximal minimum distances to previous ones~\cite{berger2016cite2vec}. \textsc{Z-order}-based sampling uses space-filling curves to sample~\cite{hu2019data}.

\uline{Advantages \& Limitations:} This category of method maintains spatial distribution and separation, which helps in identify underlying clustering patterns. However, the algorithms are time-intensive, and sampling computation time increases with the number of data points, e.g., \textsc{Blue Noise} sampling (see \autoref{fig:timeanalysis}).

\subsection{Visual Encoding} 
\label{sec-label:encoding}

\begin{wrapfigure}[3]{L}{0.09\linewidth} \vspace{-11pt}
    \begin{minipage}[t]{1.2\linewidth}
        \includegraphics[width=33pt]{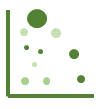}
    \end{minipage}
\end{wrapfigure}
Once data are subsampled, they are rendered multiple times, varying several visual encodings. Prior studies have demonstrated the effect of visual encodings on analysis tasks~\cite{szafir2018modeling, gramazio2014relation, cleveland1984graphical}, and visual encodings influence group or separation perception~\cite{wong2010points}, such as, color, size, shape~\cite{sedlmair2012taxonomy}, orientation~\cite{cohen2008perceptual}, texture~\cite{anobile2016number}, opacity \cite{micallef2017towards}, density~\cite{wilkinson2005graph}, motion and animation~\cite{etemadpour2017density,veras2019saliency,chen2018using}, chart size~\cite{heer2009sizing}, and others. Additionally, studies have demonstrated a perceptual effect in scatterplots with changes in the factors, including data distribution types, number of points, the proximity of concentrations of points, data point opacity, and relative density~\cite{gramazio2014relation, chung2016ordered, kim2018assessing, szafir2018modeling, gleicher2013perception, sadahiro1997cluster,correll2018looks}. Visual encoding is dependent on point size (PS) and point opacity~(OP). By default, users are provided a pre-selected set of values for these parameters (see \autoref{sec-label:model}), but they may limit them to a subset, if desired.

\subsection{Threshold Plot: Computation of Saliency Score}
\label{sec-label:threshold}

\begin{wrapfigure}[3]{L}{0.09\linewidth} \vspace{-11pt} \begin{minipage}[t]{1.2\linewidth}
        \includegraphics[width=33pt]{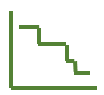}
   \end{minipage}
\end{wrapfigure} 
Next, we take the generated scatterplots and compute \textit{threshold plots} and a saliency score. The threshold plot is a monotonic step function, where the horizontal axis encodes values that describe the separation of clusters, while the vertical axis describes the number of clusters visible at that threshold. We extract from this plot the number of clusters an individual is likely to see and exactly how salient those clusters are.

\subsubsection{Merge Tree Model of Visual Density}    
We utilize the visual density-based model, first introduced by Quadri and Rosen~\cite{quadri2020modeling}, which attempts to directly identify the \textit{relative} visual density, i.e., the number of filled pixels, at which users will differentiate between clusters. They showed that this visual density-based model was a good proxy for predicting the number of clusters a human would perceive in a scatterplot. In contrast, for this paper we are trying to show that this same model can be used for design optimizations of scatterplots. We briefly summarize their approach. 

The model first encodes the clustering structure as a function of density using a merge tree. The merge tree is a data structure from Topological Data Analysis that encodes the merging order of sublevel sets of the visual density. As shown in \autoref{fig:Thresholdplot_generation}, the basic process is: (a)~an input scatterplot \textit{image} (integrating all the design factors, including SA, SR, PS, and OP) has its (b)~density histogram calculated at a pre-determined size of $20\times20$, as proposed by~\cite{quadri2020modeling}. (c)~For a given density value, $t$, \textit{x} number of clusters are observed using the 8-connected neighbors. (d)~The merge tree tracks the appearance and merging of clusters (i.e., when clusters blend to be perceived as one) across all density values, $t:0\rightarrow\infty$. The merge tree is efficiently calculated using the join tree of a scalar field (see~\cite{rosen2018hybrid} for an efficient algorithm).

The next step is that for each cluster identified in the merge tree, the persistence~\cite{zomorodian2005computing} of that cluster, $\rho$, is calculated by considering the difference between the highest ($t_h$) and lowest ($t_l$) density values where that cluster is visible, i.e., $\rho=t_h-t_l$. The fundamental intuition behind persistence is that it measures the relative scale of a feature (e.g., the relative change in density), as opposed to the absolute scale of the feature (e.g., the absolute density value). (e)~The threshold plot encodes for a given threshold, exactly how many clusters have a persistence greater than or equal to it.

\begin{figure}[!t]
    \centering
    \includegraphics[width=0.995\linewidth]{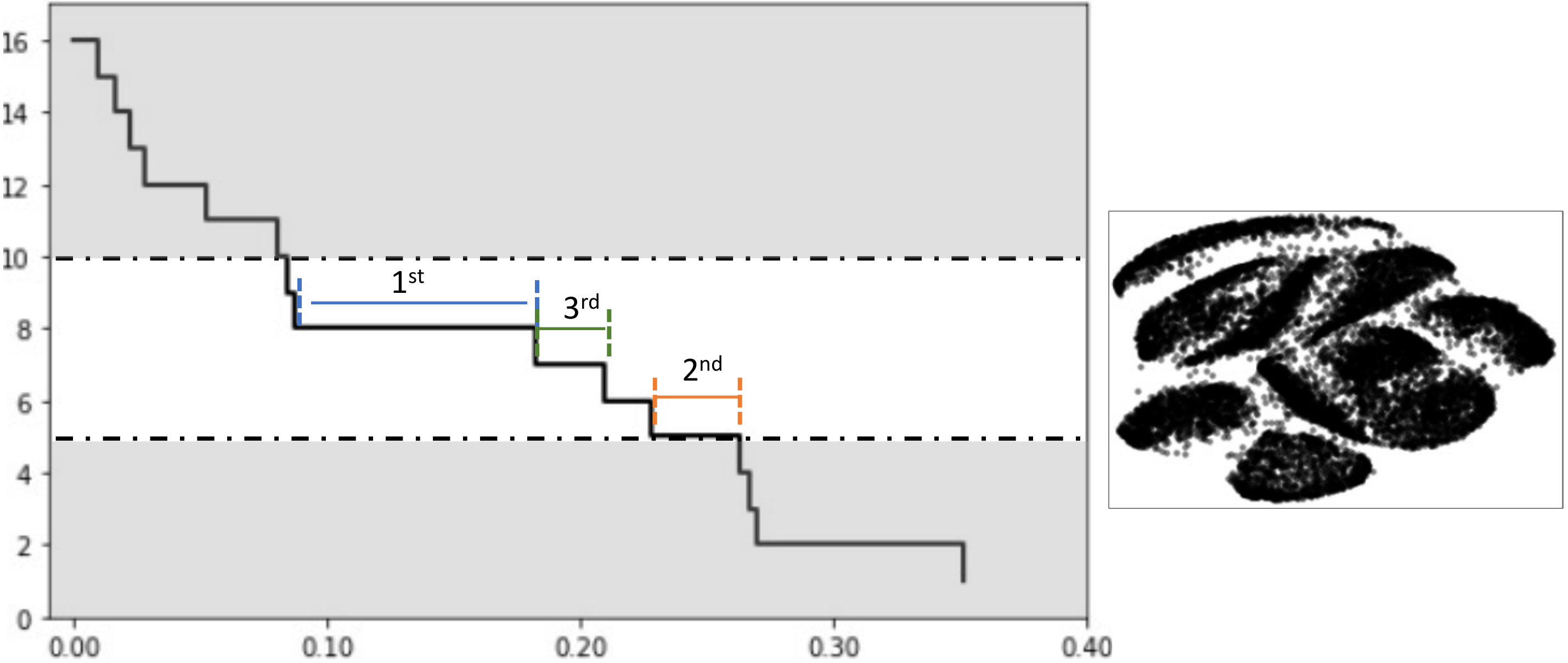}
    
    \caption{The threshold plot shows the persistence on the x-axis and number of clusters on the y-axis. The longer the bar (i.e., the saliency), the more visible the clustering structure. Within a user-selected range (white ribbon) of 5-10 clusters, the three prominent bars \textcolor{blue}{\textbf{-----}} \textcolor{orange}{\textbf{---}} \textcolor{green}{\textbf{--}} are highlighted, but \textcolor{blue}{\textbf{-----}} is the most salient.}
    \label{fig:thresholdplot_A} 
\end{figure}

\subsubsection{Scatterplot Cluster Saliency}
Using the threshold plots as-is would represent an under-constrained optimization, as it would require, at the very least, a user specification of the number of clusters or a persistence threshold.

Therefore, we wish to optimize by maximizing the \textit{dynamic range} or \textit{saliency} in the threshold plot. For a given number of clusters, $C_i$, the saliency is $T_{i,max}-T_{i,min}$ (see \autoref{fig:Thresholdplot_generation}(e)). We represent the saliency of the scatterplot by the length of the bar with the maximum individual saliency. In other words, the largest saliency is considered the best representation of the clustering structure of the scatterplot. By default all cluster counts are consider. However, users may also limit which bars are considered by selected a range, $[C_{min},C_{max}]$, for the number of clusters of interest. In \autoref{fig:thresholdplot_A} there are three prominent threshold bars \textcolor{blue}{\textbf{-----}} \textcolor{orange}{\textbf{---}} \textcolor{green}{\textbf{--}}, in the range of $5-10$ clusters, but the bar \textcolor{blue}{\textbf{-----}} represents the most salient clustering structure in the scatterplot.

\begin{figure}[!b]
    \centering
    \includegraphics[width=0.995\linewidth]{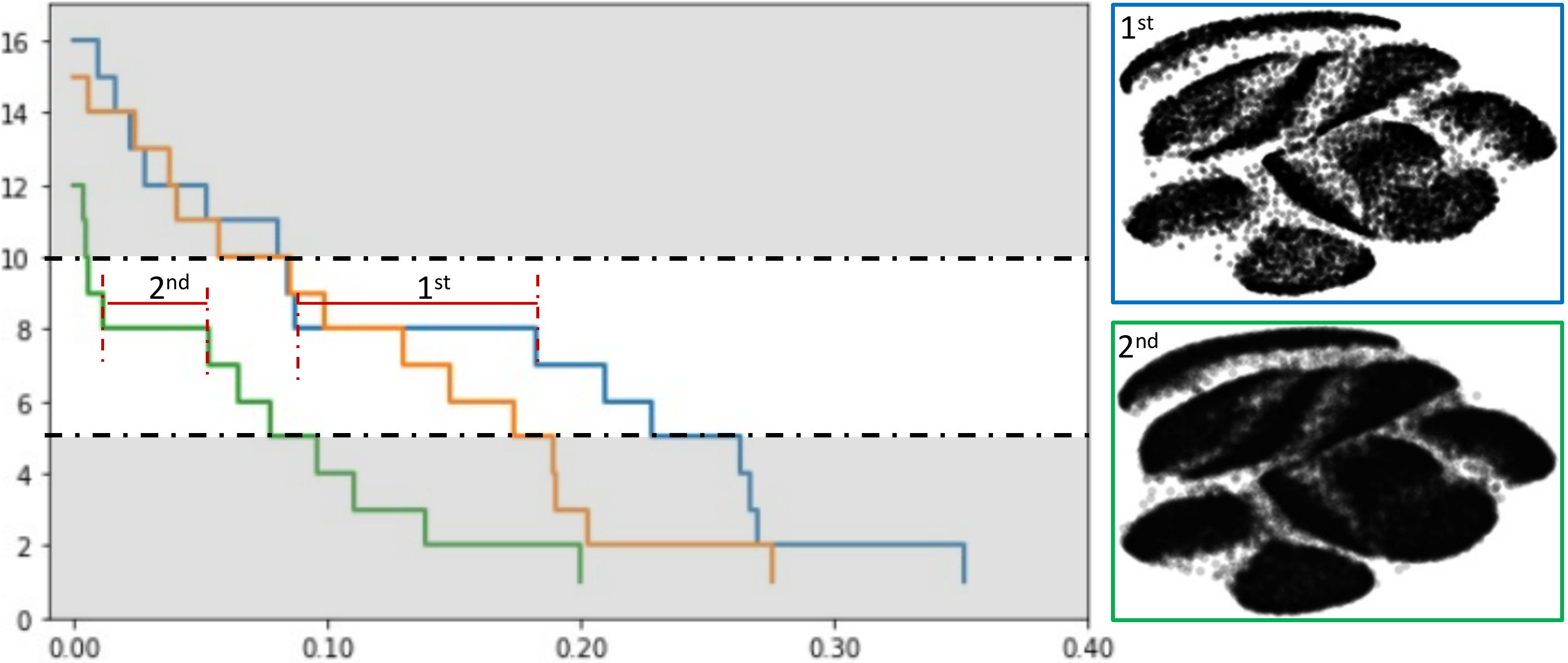}
    
    \caption{By ranking the saliency of scatterplots, we identify the one with the clearest cluster structure. In this example, there are two prominent bars within the user-defined range of 5-10 clusters (white ribbon) \textcolor{blue}{\textbf{-----}} \textcolor{green}{\textbf{--}}, but \textcolor{blue}{\textbf{-----}} represents the clearer cluster structure.}
    \label{fig:thresholdplot_B}
\end{figure}

\subsection{Optimized Design}
\label{sec:optimize}

\begin{wrapfigure}[3]{L}{0.09\linewidth}         \vspace{-11pt}
\begin{minipage}[t]{1.2\linewidth}
        \includegraphics[width=33pt]{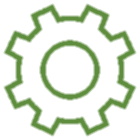}
        \end{minipage}
\end{wrapfigure} To enable selecting the best design, the saliency (i.e., maximum bar lengths) are used to directly compare scatterplots. In \autoref{fig:thresholdplot_B}, the bar length of the blue scatterplot indicates that it has more salient clustering structure that the scatterplot in green, a quality that can also be observed in the scatterplots themselves.

Finding the optimal scatterplot is done by first rendering all combination of 
SA, SR, PS, and OP (from their minimum to maximum user defined ranges). The threshold plots and scatterplot saliency are calculated.
The optimized scatterplot design is ranked using saliency score by selecting:
        \begin{itemize}
                \item SA among the finite set of sampling algorithms in \autoref{sec-label:sampling};
                \item SR among the finite set $\{SR_{min}, ..., SR_{max}\}$;
                \item PS among the finite set $\{PS_{min}, ..., PS_{max}\}$;
                \item OP among the finite set $\{OP_{min}, ..., OP_{max}\}$; and
                \item $C_{min}$ and $C_{max}$ are user-selected but not optimized.
        \end{itemize}
Finally, all scatterplots are ranked from most salient to least salient and provided to the user.




\section{User-Guided Optimization Interface}
\label{sec-label:model}

We developed an interactive web interface to demonstrate our approach (which is used in \autoref{sec-label:casestudy}), as seen in \autoref{fig:interactivemodel}, where one can select an optimized scatterplot based on the cluster structure saliency. The interface enables optimizing visual encodings, i.e., point size (PS) and opacity (OP), and data aspects, i.e., subsampling algorithm (SA) and sampling rate (SR), on real-world data using the saliency of threshold plots. The user can select the ranges for parameters ($PS_{min},PS_{max},OP_{min},OP_{max},SR_{min},SR_{max}$) and  cluster count ($C_{min},C_{max}$) for more refined ranking. Here, we detail the stages illustrated in the overview from \autoref{fig:overview}.

\para{Operation of the Interface} Our interface outputs the ranking of scatterplots by their cluster structure saliency, also known as \textit{saliency score}. The user selects the dataset and can optionally choose different ranges for sampling rate, point size, opacity, and cluster count. The output ranks and presents the scatterplots with the highest saliency.

\para{Input Datasets} We selected datasets from the prior studies in visualization as our experimental data. We selected eight representative datasets (see \autoref{fig:casestudy_images}) with different characteristics: six datasets with clustering structures (MNIST~($n=70000$)~\cite{lecun1998gradient}, Conditional Based Maintenance~($n=10000$)~\cite{coraddu2016machine}, Clothes~($n=26569$)~\cite{xia2017ldsscanner}, Crowdsourced Mapping~($n=10845$)~\cite{johnson2016integrating}, Epileptic Seizure~($n=11500$)~\cite{andrzejak2001indications}, Swiss Roll~2D~($n=8000$)~\cite{sedlmair2013empirical} and two with curved stripes (Swiss Roll~3D~($n=10000$)~\cite{sedlmair2013empirical} and Abalone~($n=4177$)~\cite{dua2017uci}). Four additional examples appear only in our demo application: census-income~($n=48842$)~\cite{kohavi1996scaling},
pp-gas-emission-2011~($n=36733$)~\cite{kaya2019predicting}, 
credit-card~($n=30000$)~\cite{yeh2009comparisons}, 
diabetes-data~($n=50000$, only half of the dataset)~\cite{strack2014impact}.
For datasets with dimensionality higher than two, we first transformed them into 2D data using t-SNE and normalized them to $[0,1]\times[0,1]$.

\begin{wrapfigure}[3]{L}{0.09\linewidth}         
\begin{minipage}[t]{1.2\linewidth}
        \includegraphics[width=33pt]{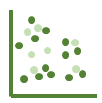}
        \end{minipage}
\end{wrapfigure} 
\noindent
\para{Subsampling (SA and SR)} We subsample the dataset using the 14~algorithms (SA) from \autoref{tbl:sampling}. To understand the performance of each sampling technique, we selected sampling rates (SR) on the interval $[5\%,95\%]$ of the input data with a step size of 5\%.

\begin{wrapfigure}[3]{L}{0.09\linewidth}         \vspace{-11pt}
\begin{minipage}[t]{1.2\linewidth}
        \includegraphics[width=33pt]{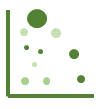}
        \end{minipage}
    \end{wrapfigure}
    \noindent
\para{Visual Encoding (PS and OP)} Data are presented as point marks (i.e., circles~$\bullet$) on the scatterplot, and two visual encodings that influence visual density are varied. The point size (PS) is selected to have area $\{20_{px}, 40_{px}, 60_{px}, 80_{px}\}$, and the point opacity (OP) is chosen to be $\{1\%, 5\%, 10\%, 20\%, 40\%, 80\%\}$. Both ranges and step sizes are selected using guidelines from~\cite{quadri2020modeling}.

\para{Scatterplot Rendering} For each dataset, scatterplots were rendered using all combinations of $SA \times SR \times PS \times OP$. By default, they are rendered with image dimension ($[X \times Y]$) $[700_{px} \times 700_{px}]$, which was selected such that the image would fit vertically on the majority of desktop monitors without scrolling~\cite{StatCounter2019} with a horizontal resolution to match. Any data falling outside this region is clipped.

\begin{wrapfigure}[3]{L}{0.09\linewidth}         \vspace{-11pt}         \begin{minipage}[t]{1.2\linewidth}
        \includegraphics[width=33pt]{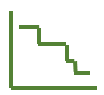}
    \end{minipage}
\end{wrapfigure} 
\noindent
\para{Saliency Computation} The fundamental unit our approach is the \textit{visual density}, in particular the point at which human perception of the visual density of cluster distributions will blend to be perceived as one. For each scatterplot generated in the prior steps, a threshold plot is generated, and the cluster structure saliency of the scatterplot is calculated.

\begin{wrapfigure}[3]{L}{0.09\linewidth}         
\vspace{-11pt} 
\begin{minipage}[t]{1.2\linewidth}
        \includegraphics[width=33pt]{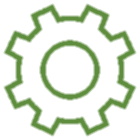}
    \end{minipage}
\end{wrapfigure} 
\noindent
\para{Optimized Design} The final results are the ranked order of scatterplot designs based on their saliency. One point to be noted here is that many scatterplot designs produce similar saliency values because they are perceptually similar (refer to \autoref{sec-label:userstudy} for more details).

\section{Quantitative Analyses}
\label{sec-label:computation}

To understand the operational aspects of our approach, we performed the following analyses on eight datasets.

\subsection{Computation Time}

\begin{figure}[!b]
    \includegraphics[width=0.995\linewidth]{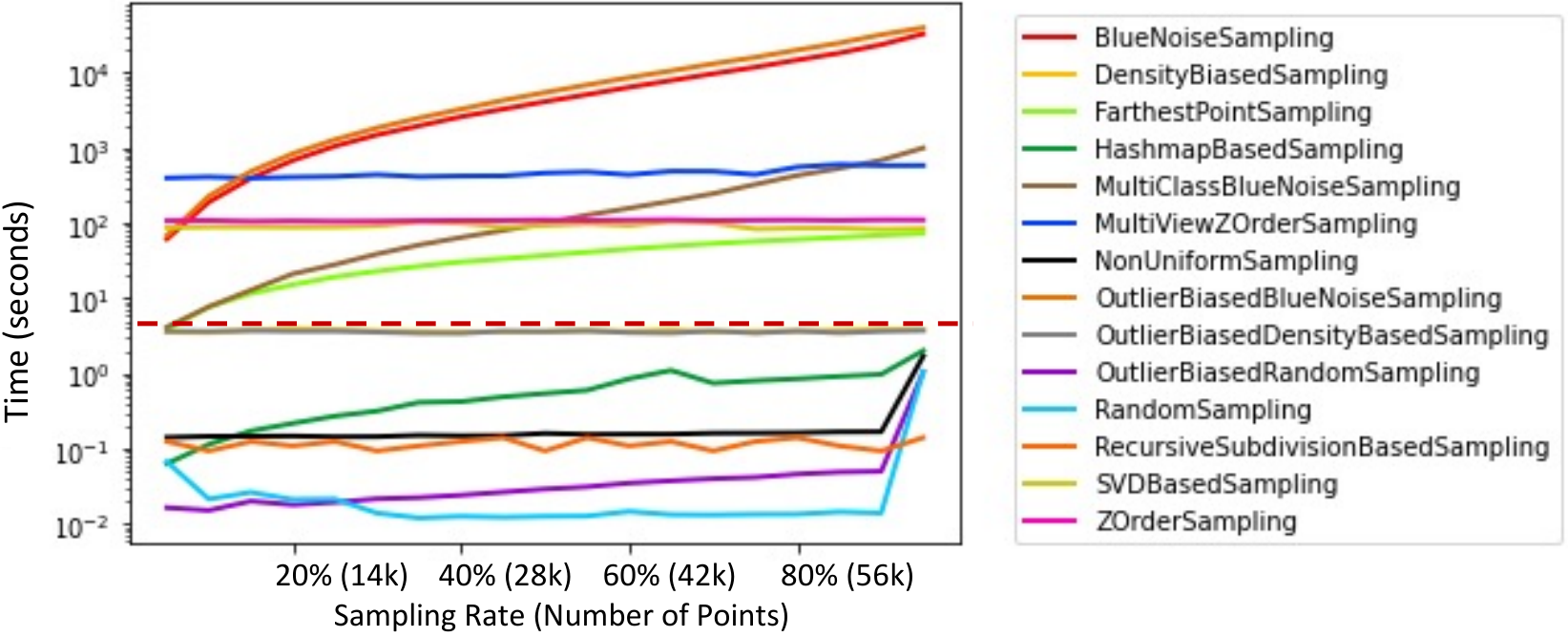}
        \caption{Subsampling computation time (logarithmic scale) for 14 algorithms across different sampling rates (linear scale, 5\% to 95\% with interval of 5\%) for the MNIST dataset. The dashed red line \textcolor{red}{\textbf{- - -}} represents the time needed to render an image (see \autoref{fig:compute_time}).}
        \label{fig:timeanalysis}
\end{figure}

\para{Subsampling} We recorded computation time for subsampling on every dataset, on each algorithm, and for each sampling rate. We observed similar patterns for all datasets. Therefore, we will discuss the results for only MNIST. \autoref{fig:timeanalysis} shows the results. The approaches roughly break down into three groups with low (e.g., \textsc{Random} and \textsc{Outlier Biased} sampling), medium (e.g., \textsc{SVD-based} and \textsc{Farthest point} sampling), and high (e.g., \textsc{Blue noise} and \textsc{Outlier Biased Blue Noise} sampling) completion time. The second observation is that some algorithms (e.g., \textsc{Non-uniform}, \textsc{Outlier Biased Density}, and \textsc{Random} sampling) perform uniformly for all sampling rates, while other algorithms (e.g., \textsc{Blue Noise} and \textsc{Outlier Biased Blue Noise}) have completion times that increase as the sampling rate increases.

\begin{figure}[!t]
    \includegraphics[width=0.995\linewidth]{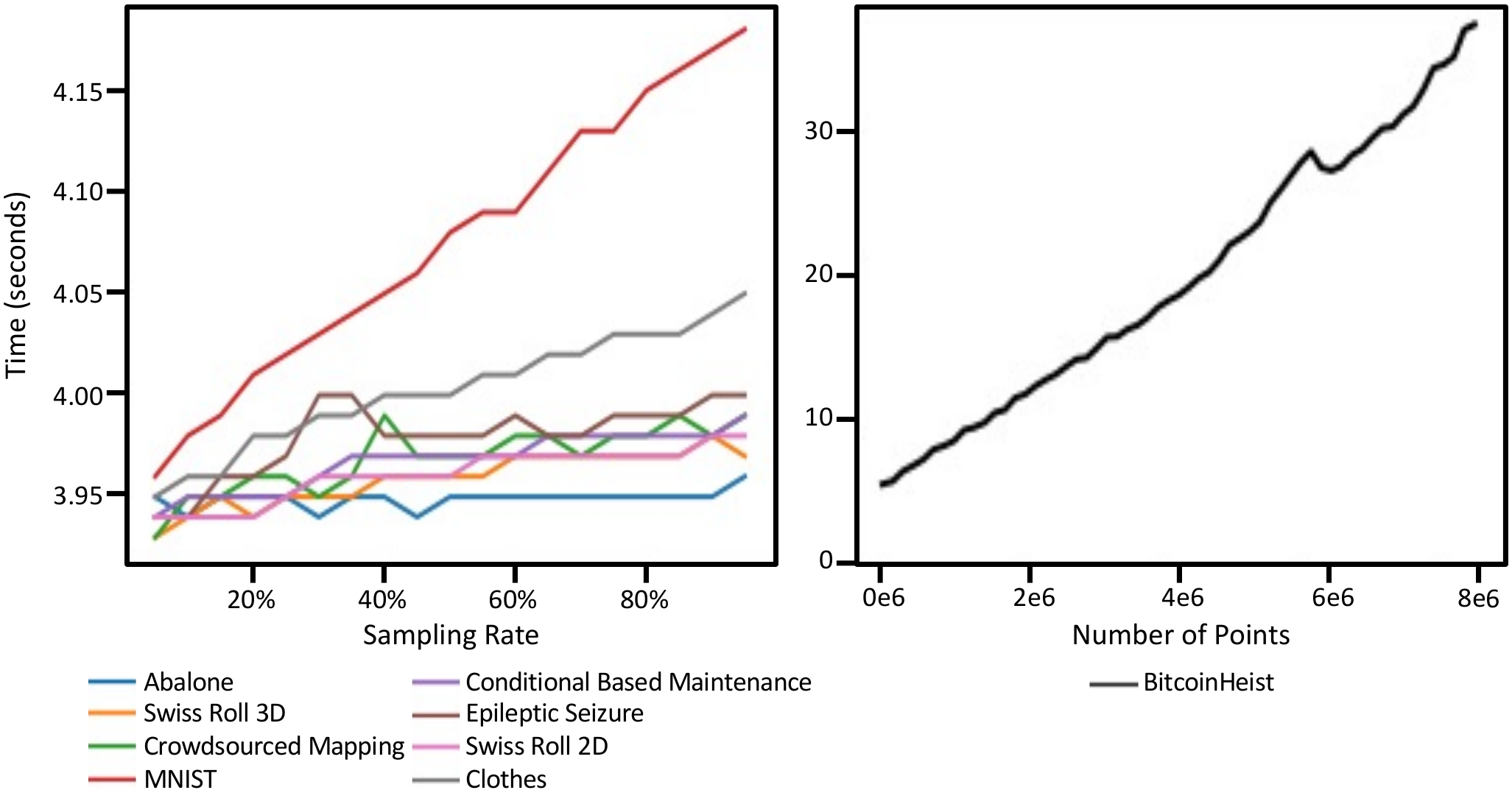}
    
    \vspace{-15pt}
    \hspace{0.05\linewidth}
    \subfloat[8 datasets\label{fig:timeanalysis_model}]{\hspace{0.46\linewidth}}
    \subfloat[BitcoinHeist \label{fig:bitcoin}]{\hspace{0.46\linewidth}}
    
    \caption{Rendering and saliency computation time for (a)~8~datasets used in all analysis and (b)~\textit{BitcoinHeist}~\cite{akcora2019bitcoinheist} dataset used for scalability. Each point is the average time for 14 (sampling algorithms) $\times$ 4 (point sizes) $\times$ 5 (point opacities). Sampling computation time  (linear scale) is dependent on number of points  (linear scale), i.e., the sampling rate.}
    \label{fig:compute_time}
\end{figure}

\para{Rendering and Saliency Extraction} After subsampling the data, the scatterplot is rendered, the threshold plot is calculated, and the saliency is extracted. We computed and recorded the average time taken for each dataset (excluding subsampling) in \autoref{fig:timeanalysis_model}. The computation time fits in a relatively small window around 4 seconds, though a general trend shows that the computation time is proportional to the number of data points, probably owed to increased rendering costs. Furthermore, the dashed red line in \autoref{fig:timeanalysis} shows the average time for the rendering and saliency computation time of approximately 4 seconds. While several subsampling methods take less time, many require significantly more. Hence, there is a trade off between time and quality, which we explore in the next section.

\para{Scalability} To further analyze the computation for more data points in terms of scalability, we selected a dataset, \textit{BitcoinHeist}~\cite{akcora2019bitcoinheist}, with approximately 3 million data points. We computed and recorded the computation time for rendering and saliency calculations. The trend seen in \autoref{fig:bitcoin} demonstrates the linear characteristic of computation time with number of data points.

\subsection{Subsampling Quality} 
The computation of subsampling is a significant portion of processing time. An important question to reflect upon is whether all of the subsampling methods are necessary, particularly those requiring high computation time, e.g., in \autoref{fig:timeanalysis}, \textsc{Blue Noise} and \textsc{Outlier Biased Blue Noise} sampling take several orders of magnitude more compute time compared to \textsc{Random} sampling or \textsc{Outlier Biased Random} sampling.

\begin{figure}[!t]
    \includegraphics[width=0.995\linewidth]{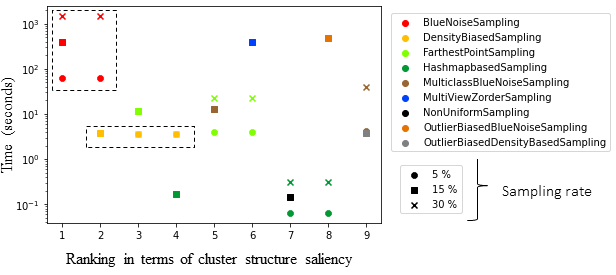}
    \caption{Evaluation of SA time computation vs.\ performance on three sampling rates (5\%, 15\%, and 30\%) on the MNIST dataset. The shape marker represents the SR, and the color represents the SA. The chart shows computation time and the rank of the top nine sampling methods. The top left corner shows \textsc{Blue Noise} sampling as the top ranked method, but it also required more computation time.}
    \label{fig:samplingvstime}
\end{figure}

We performed a comparative analysis of the algorithms by selecting the MNIST dataset with a sampling rate between 5\% to 30\% and looked at the top-ranked methods. \autoref{fig:samplingvstime} shows the evaluation of 14 SA time computations against their performance by measuring how frequently the SA produces the optimal scatterplot design. We included top nine SAs at three pre-selected SRs in the figure.

We found that \textsc{Blue Noise} and \textsc{Density Biased} sampling methods are the top two ranked algorithms, followed by \textsc{Farthest Point} sampling as the third in line (see \autoref{fig:samplingvstime}). The main reason behind this ranking is the feature preservation, i.e., spatial separation (see \autoref{tbl:sampling}). From the top two methods, we found that there is a higher computation time for \textsc{Blue Noise} than \textsc{Density Biased} sampling methods. However, \textsc{Blue Noise} generated more salient structure, while \textsc{Density Biased} produced slightly less salient structure but also took less time comparatively. The important conclusion here is that some techniques that take longer to compute can provide the best results.

\section{User Study}
\label{sec-label:userstudy}

To validate the utility of the threshold plot saliency for ranking scatterplots based on their clustering structure, we ran a user study on Amazon Mechanical Turk (AMT).

\subsection{Study Design}

\subsubsection{Hypotheses}
    
\begin{description}[leftmargin=!,labelindent=5pt,itemindent=-20pt]

    \item[{[H1]}\label{Hyp:H1}] \textit{Similar patterned threshold plots represent scatterplots that are perceptually similar and have similar cluster structures.}

\end{description} 

We believe that threshold plots can be used as a proxy to identify which scatterplot designs have more salient structure, and scatterplots with similar threshold plot shapes have similar visual density and visual separation.

\begin{description}[leftmargin=!,labelindent=5pt,itemindent=-20pt]

    \item[{[H2]}\label{Hyp:H2}]  \textit{The longer the maximum threshold bar, the more salient the cluster structure is in a scatterplot.}
    
\end{description} 

We further consider the length of the longest bar, i.e., \textit{saliency}, to be a valid feature for ranking scatterplot designs.

\subsubsection{Study Task} 
We utilize two tasks, \ref{Task:T1} and \ref{Task:T2}, for \ref{Hyp:H1} and \ref{Hyp:H2}, respectively.

\begin{description}[leftmargin=!,labelindent=5pt,itemindent=-20pt]
    \item[{[T1]}\label{Task:T1}] \textit{Which scatterplot has more similar cluster structure to the reference scatterplot?}
\end{description} 

     A reference and two other scatterplot designs are shown, and subjects have to select the scatterplot design with the more similar cluster structure to the reference plot.

\begin{description}[leftmargin=!,labelindent=5pt,itemindent=-20pt]

    \item[{[T2]}\label{Task:T2}] \textit{Which scatterplot has a clearer cluster structure?}
\end{description}     

    Two scatterplots are shown, and subjects have to select the one with a clearer cluster structure. Each scatterplot has a calculated saliency value, and those with a higher saliency value should have a clearer cluster structure.

\subsection{Stimulus Generation}

\para{Data Selection} We selected six datasets (MNIST, Conditional Based Maintenance, Clothes, Epileptic Seizure, Swiss Roll 2D, and Swiss Roll 3D) from those listed in \autoref{sec-label:model}. In addition,  Crowdsourced Mapping was used for the training examples, and Abalone was excluded for having a similar shape in the scatterplot to the Swiss Roll 3D.

\para{Scatterplot Rendering} The scatterplot images are rendered with similar parameters as those in the interface.
\begin{itemize}[noitemsep,itemsep=4pt]
 
    \item Stimuli dimensions ($[X \times Y]$): $[700_{px} \times 700_{px}]$
        
    \item Data point size/area ($PS$): $\{20_{px}, 40_{px}, 60_{px}, 80_{px}\}$ 

    \item Data point opacity ($OP$): $\{1\%, 5\%, 10\%, 20\%, 40\%, 80\%\}$
         
    \item Sampling rate (as a proportion of the number of data points) ($SR$): On the interval $[5\%,95\%]$ with $5\%$ step size using all 14 SA techniques from \autoref{tbl:sampling}.
 
\end{itemize}

\subsection{Study Procedure}
\label{sec-label:study_procedure}

\subsubsection{\ref{Hyp:H1} Threshold Plot Difference as Perceptual Similarity}
\label{sec-label:study_procedure:h1}
\begin{figure}[!b]
    \centering
    \includegraphics[width=0.975\linewidth]{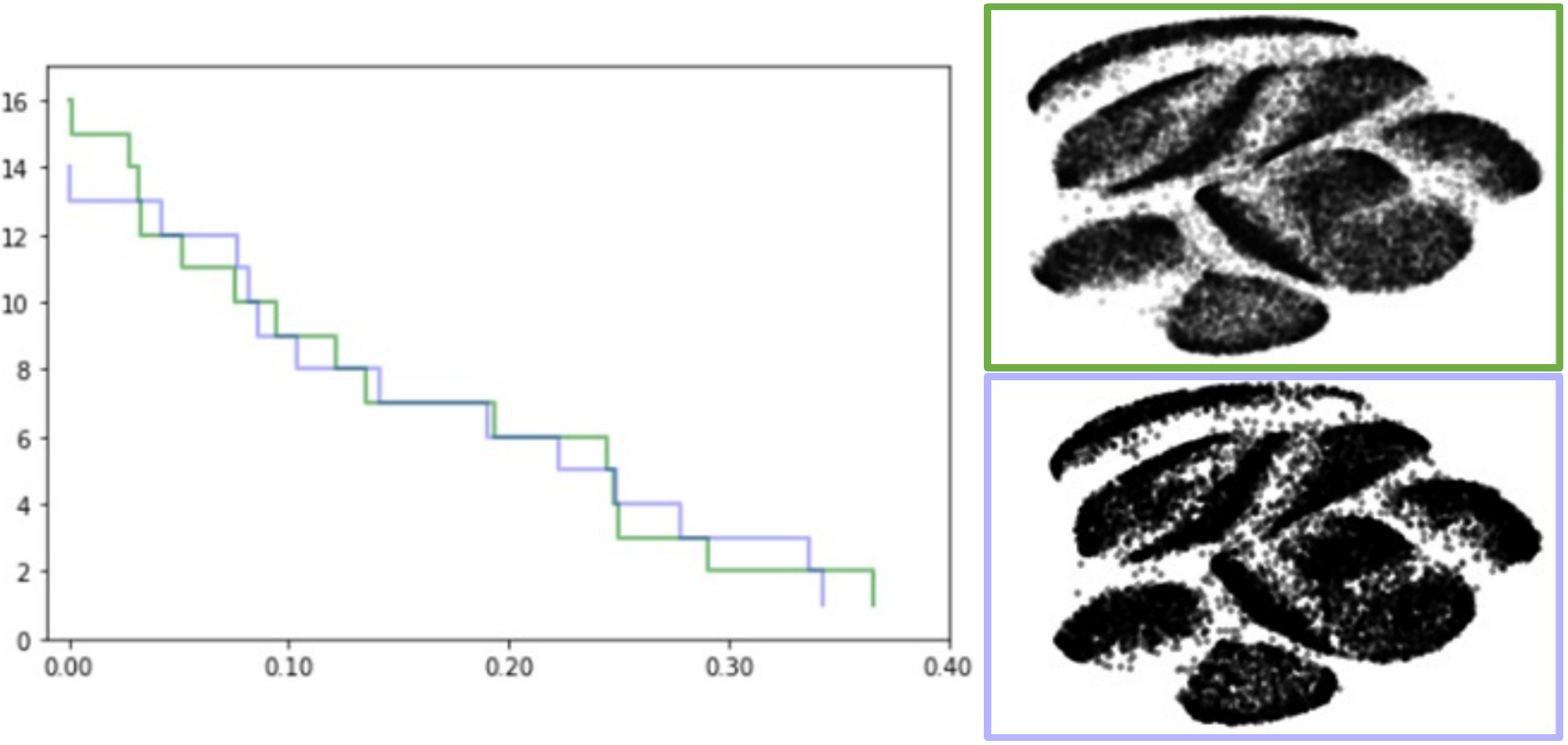}

    \caption{The threshold plots (left) show similar patterns in the clustering structure of their associated scatterplots (right).}
    \label{fig:AUC}
    
\end{figure}

Two similar scatterplots potentially represent similar cluster structures, and it can become ambiguous to distinguish between them. In our approach, two scatterplots are \textit{perceptually similar} if their threshold plots are similar to each other (e.g., see \autoref{fig:AUC}). 
As a metric to determine the perceptual similarity between clustering structures, we use the area under the curve (AUC), i.e., $AUC(X)=\sum _{i=1}^{n}\left|x_{i}\right|$ for a threshold plot, as a measure to compare between scatterplots.

\begin{figure}[!t]
    \centering
    \includegraphics[width=0.975\linewidth]{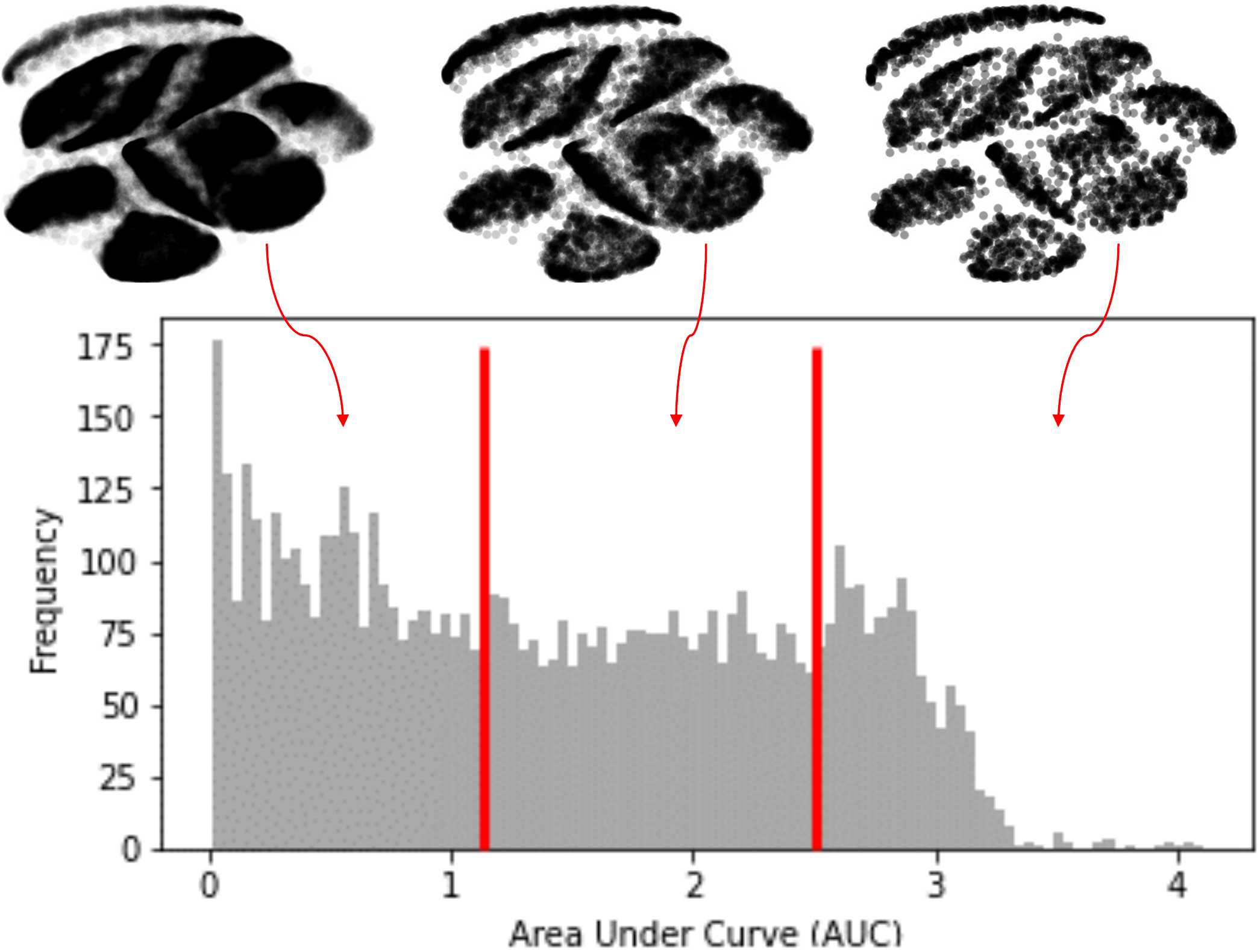}
    \caption{Histogram of Area Under Curve (AUC) for MNIST with example scatterplots.}
    \label{fig:AUC_SCP_example}
\end{figure}

We calculated the distribution of AUCs for threshold plots in the MNIST datasets (see \autoref{fig:AUC_SCP_example}). The AUCs are divided into three equally sized bins.
Finally, we use three similarity criteria: Similar (SR) if scatterplots are from the same bin; Somewhat Similar~(SS) if scatterplots are from adjacent bins, e.g., 1st and 2nd, or 2nd and 3rd; and Dissimilar~(DS) if scatterplots are from the 1st and 3rd bin.

\subsubsection{\ref{Hyp:H2} Threshold Plot Bar Length as Saliency}
\label{thresholdbar_saliency}

\begin{figure}[!b]
   \centering
    \includegraphics[width=0.995\linewidth]{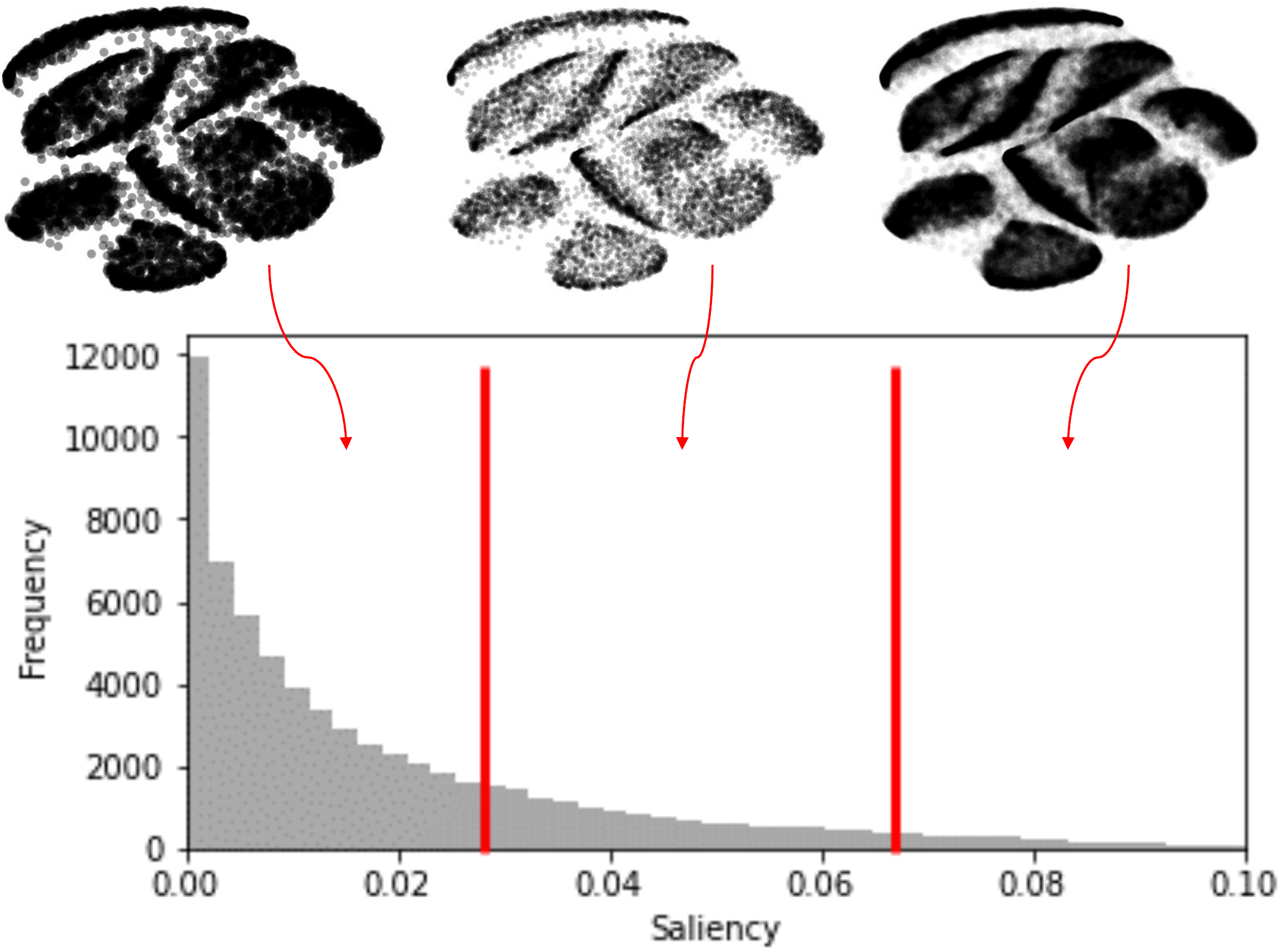}
    \caption{Histogram of saliency for all MNIST scatterplots with example scatterplots.}
    \label{fig:saliency_distribution}
\end{figure}

As we consider optimizing the saliency of plots, one question naturally occurs, which is how are the saliency values distributed across the parameters we have selected. While the precise distribution is data-dependent, all fall into a similar trend that can be observed for the MNIST dataset in \autoref{fig:saliency_distribution}. The vast majority of configurations lead to low cluster saliency, and few configurations provide the optimal saliency, which makes finding that optimal saliency by a manual search (i.e., manually selecting parameters), instead of our approach, difficult. 

For our analysis, we divide the space of saliency values for a given dataset into three evenly spaced groups: low, medium, and high saliency. In the example of \autoref{fig:saliency_distribution}, the bins are: low $[0.0,0.033)$, medium $[0.033,0.067)$, and high $[0.067,0.1]$.

\subsubsection{Stimulus and Trials}  
\label{threshold_AUC_combination}

We keep the number of trials small ($6 \times \ref{Task:T1}$ and $12 \times \ref{Task:T2}$) for both tasks to reduce the risk of learning effects. The scatterplots for stimuli are selected from the generated pool (number of datasets (D) $\times$ SA $\times$ SR $\times$ PS $\times$ OP) in random order but in the following manner:

For \textbf{\ref{Task:T1}}, the subject is shown three scatterplots in one stimulus: one reference~(R) and two as a forced-choice. The reference~(R) is displayed above and two options are shown below. From the three bins~(B) of AUC perceptual similarity scatterplots are classified as: similar (SR), somewhat similar (SS), and dissimilar (DS), with respect to the reference scatterplot, R. For two scatterplots, A and B, the possible options are: 
  \begin{enumerate}
    \item A is similar to R. B is dissimilar to R. 
    \item A is similar to R. B is somewhat similar to R.
    \item A is somewhat similar to R. B is dissimilar to R.
  \end{enumerate}
Next, we have these three combinations for the \ref{Task:T1}: $SR \times R \times DS$, $SR \times R \times SS$, and $SS \times R \times DS$ . Each trial randomly selects one combination from the above options for each dataset: $D(6) \times B(1) = 6$ stimuli for \ref{Task:T1}.

For \textbf{\ref{Task:T2}}, the subject is shown two scatterplots for one dataset as a forced-choice with saliency values divided into three bins~(B) of saliency, High~(H), Medium~(M), and Low~(L). Next, we have these six combinations for the \ref{Task:T2}: $H \times H$, $H \times M$, $H \times L$, $M \times L$, $M \times M$, $L \times L$. Each trial randomly selects two combinations among those six for each dataset: $D(6) \times B(2)= 12$ stimuli for \ref{Task:T2}.

\subsubsection{Interface} 
We used a webpage for the experiments, where each participant was given 18 stimuli (6$\times$\ref{Task:T1} and 12$\times$\ref{Task:T2}) selected randomly from the generated pool. The maximum allocated time, which was visible to participants, for one trial of \ref{Task:T1} was 15 seconds, and \ref{Task:T2} was 10 seconds. At the expiration of time, the page was automatically advanced. At the beginning of the experiment, we included a brief introduction, examples, and one training task per task type using the Crowdsourced Mapping dataset. We also included a open ended post-test questionnaire for general feedback on usability and quality. The experiment was expected to last less than 10~minutes in total. The consent was obtained at the beginning of the experiment. Please refer to the user study demo at {\small\textless\textcolor{blue}{\url{http://scatter.projects.jadorno.com/}}{\small\textgreater}}.

\subsubsection{Participants} 
We recruited participants from Amazon Mechanical Turk (AMT) for the IRB-approved study. Based upon a post hoc power analysis of the preliminary experiment data, we recruited a total of $70$ participants (49 male, 21 female; ages: $[18-64]$, median age group: $[25-44]$) limited to the US or Canada. 47\% of participants reported having corrected vision. All participants had a HIT approval rate of $\geq 95\%$ and were compensated at US Federal minimum wage or above.

In total, task \ref{Task:T1}: $6$ trials $\times$ $70$ participants $=420$~responses, and task \ref{Task:T2}: $12$ trials $\times$ $70$ participants $=840$~responses, were collected. We carried out some data quality checks on responses with the constraints: participant responses with timing less than one second for a given trial were rejected, and trials with no response or expired time were rejected. We identified one trial for task \ref{Task:T2} with no response from the subject and hence rejected, leaving a total of $839$~responses for analysis.

\subsection{Analysis Methodology}

\subsubsection{Model Specifications}
To test our hypotheses, we fit models predicting whether subjects would choose the theoretically-predicted ("target") response option rather than the alternative ("comparison") option. This was a binary choice, so we modeled the choice using Bernoulli regression models. A Bernoulli regression is a generalized linear model that estimates the probability of an event (i.e., the subject choosing the target option) occurring using a linear combination of predictors. To ensure only valid probabilities in the [0,1] range were estimated, we used the logit (log-odds) link function. In each model, our focal predictors were the proxy indicators for the visualization characteristic hypothesized to determine choice in our theoretical model (i.e., AUC for \ref{Task:T1}, saliency/bar length for \ref{Task:T2}) for the two response options.

Support for our hypotheses would be indicated if the proxy indicators strongly and significantly predicted subject choice of the target option. When there is a large difference between proxy indicator values for the two options, subjects should select the target option with high probability. However, when the two options have similar proxy indicator values, then subjects should select the two options at similar rates (i.e., select the target option with probability $\approx .50$).

\para{Task 1} In \ref{Task:T1}, we predicted subjects would choose the scatterplot with the smaller AUC from the reference plot as being more similar in cluster structure. AUC captures how similar a plot is to the reference plot. Small AUC differences indicate that the plot is highly similar to the reference plot. Large AUC differences indicate that the plot is highly dissimilar to the reference plot. We predicted that the probability of choosing the target option should increase the larger the difference in AUC values between the two response options. That is, when the target is much more similar to the reference plot than is the comparison plot, participants should consistently choose the target plot. When the two plots are equally similar (or dissimilar) to the reference plot, participants should select each at similar rates. As described above, we modeled that probability that the subject would choose the target (more-similar) option in trial $t$ using the AUC values for the target (TAUC) and comparison (less-similar; CAUC) options and included random intercepts for each respondent ($i$) and stimulus dataset ($j$) to account for dependency across trials:

\begin{align}
ChooseTarget_{ijt} &\sim Bernoulli(p_{ijt}) \\
logit(p_{ijt}) &= \beta_{1} \times TAUC_{ijt} + \beta_{2} \times CAUC_{ijt}\ + \nonumber \\  
&\qquad \alpha + \gamma_{i} + \delta_{j} \nonumber \\
\gamma_{i} &\sim Normal(\mu_{\gamma}, \sigma_{\gamma}) \nonumber \\  
\delta_{j} &\sim Normal( \mu_{\delta}, \sigma_{\delta}) \nonumber 
\end{align}

%
%
%
%
%

\begin{table}[!t] 
\centering
    \caption{Model predictions for \ref{Task:T1} show the predicted probabilities of choosing the target option (Pr(CT)) based on the fitted model, with 95\% CI, for combinations of target and comparison AUC (see \autoref{sec-label:study_procedure:h1}) values. For SR, SS, and DS, see \autoref{threshold_AUC_combination}. SE=standard error, CI=confidence interval.}
    \label{tbl:modelpredictionsT1}

    \resizebox{0.95\linewidth}{!}{%
    \begin{tabular}{>{\raggedleft\arraybackslash}p{.09\textwidth}|>{\raggedleft\arraybackslash}p{.09\textwidth}|>{\centering\arraybackslash}p{.056\textwidth}|>{\centering\arraybackslash}p{.053\textwidth}|>{\centering\arraybackslash}p{.035\textwidth}|>{\centering\arraybackslash}p{.085\textwidth}}
\toprule
   \multicolumn{1}{c|}{Target} & \multicolumn{1}{c|}{Comp.} & AUC & & & \\ 
   \multicolumn{1}{c|}{[AUC]} & \multicolumn{1}{c|}{[AUC]} & diff. & Pr(CT) & SE & 95\% CI \\
\hline
SR {[}0.0{]}{ }{ }{ }{ } & SR {[}0.0{]}{ }{ }{ }{ } & 0.00 & 0.61 & 0.09 & {[}0.43, 0.77{]} \\
SS {[}1.0{]}{ }{ }{ }{ } & SS {[}1.0{]}{ }{ }{ }{ } & 0.00 & 0.64 & 0.08 & {[}0.48, 0.78{]} \\
DS {[}2.0{]}{ }{ }{ }{ } & DS {[}2.0{]}{ }{ }{ }{ } & 0.00 & 0.67 & 0.11 & {[}0.44, 0.84{]} \\
SR {[}0.0{]}{ }{ }{ }{ } & SS {[}1.0{]}{ }{ }{ }{ } & 1.00 & 0.75 & 0.06 & {[}0.62, 0.85{]} \\
SS {[}1.0{]}{ }{ }{ }{ } & DS {[}2.0{]}{ }{ }{ }{ } & 1.00 & 0.77 & 0.05 & {[}0.65, 0.86{]} \\
DS {[}2.0{]}{ }{ }{ }{ } & DS {[}3.0{]}{ }{ }{ }{ } & 1.00 & 0.79 & 0.08 & {[}0.60, 0.91{]} \\
SR {[}0.0{]}{ }{ }{ }{ } & DS {[}2.0{]}{ }{ }{ }{ } & 2.00 & 0.85 & 0.05 & {[}0.74, 0.92{]} \\
SS {[}1.0{]}{ }{ }{ }{ } & DS {[}3.0{]}{ }{ }{ }{ } & 2.00 & 0.86 & 0.04 & {[}0.75, 0.93{]} \\
DS {[}2.0{]}{ }{ }{ }{ } & DS {[}4.0{]}{ }{ }{ }{ } & 2.00 & 0.88 & 0.06 & {[}0.71, 0.96{]} \\
SR {[}0.0{]}{ }{ }{ }{ } & DS {[}3.0{]}{ }{ }{ }{ } & 3.00 & 0.91 & 0.04 & {[}0.81, 0.96{]} \\
SS {[}1.0{]}{ }{ }{ }{ } & DS {[}4.0{]}{ }{ }{ }{ } & 3.00 & 0.92 & 0.04 & {[}0.82, 0.97{]} \\
SR {[}0.0{]}{ }{ }{ }{ } & DS {[}4.0{]}{ }{ }{ }{ } & 4.00 & 0.95 & 0.03 & {[}0.85, 0.99{]} \\
\bottomrule
\end{tabular}}

\vspace{3pt}
{\footnotesize Pr(CT): Pr(ChooseTarget)}
\end{table}

\begin{figure}[!t]
      \centering
         {\includegraphics[width=0.95\linewidth]{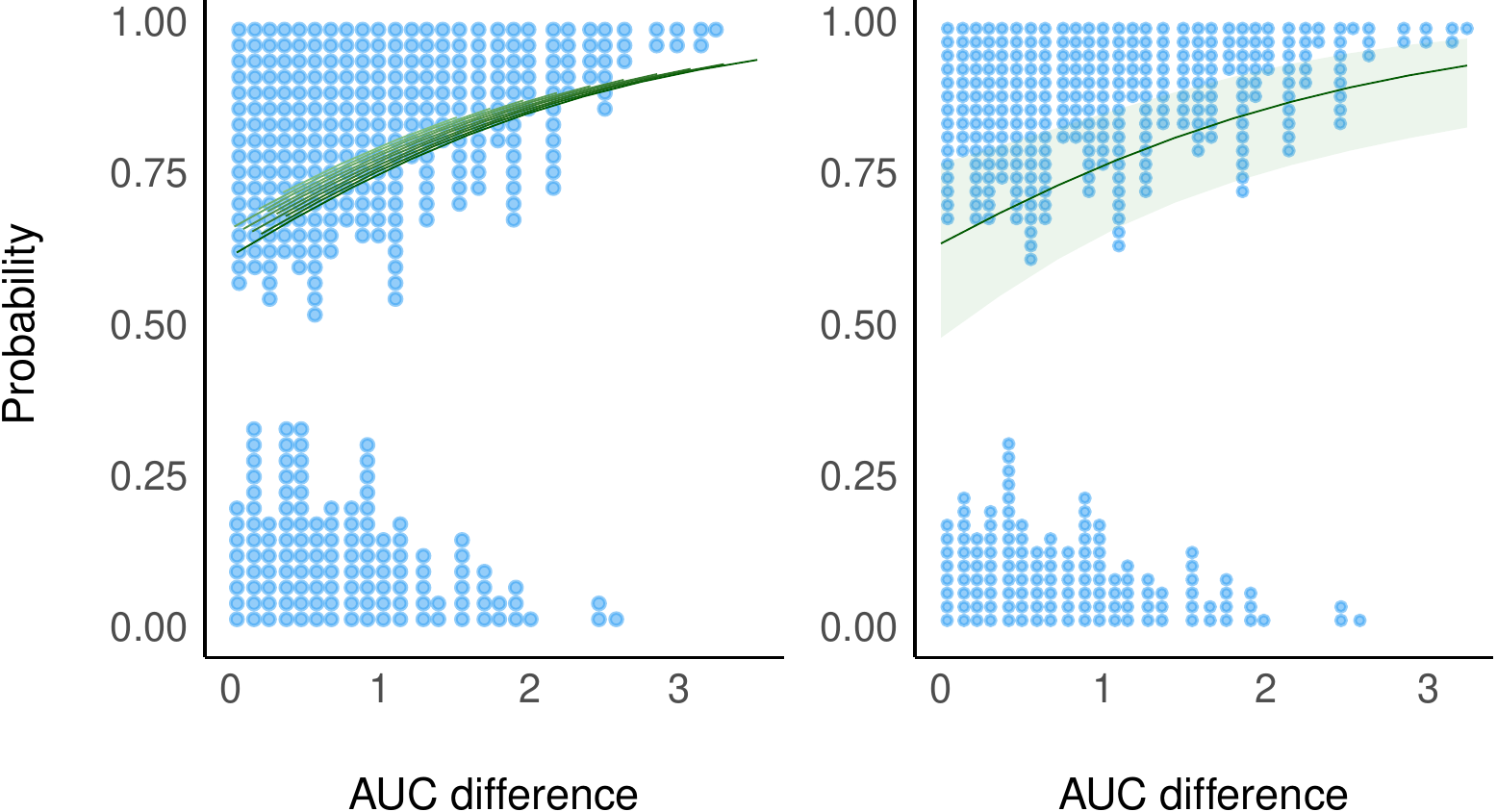}\hspace{0.025\linewidth}}

         \vspace{3pt}
         \hspace{10pt}
         \begin{minipage}[b]{.425\linewidth}
        \subcaption{Main model predictions: darker lines indicate smaller target plot AUC.}
    \end{minipage}
    \hspace{5pt}
      \begin{minipage}[b]{.425\linewidth}
        \subcaption{Predictions and 95\% CI from model with AUC difference as the predictor.}
    \end{minipage}
         
        \caption{User study results for \ref{Task:T1}, Pr(ChooseTarget) by AUC difference from reference. The blue dot represents data points for each observation with lines showing predicted probabilities of choosing the target option as a function of difference between the AUC values for the two stimuli options. 
        Panel (a) shows results for the main model, where each line reflects a combination of Target AUC and Comp.\ AUC as shown in \autoref{tbl:modelpredictionsT1}. Darker lines indicate higher Target AUC. The results show that as the AUC difference increases, probability of choosing the target option increases strongly, and there is little variation in predicted probability of choosing the target option by levels of Target AUC (different lines). Panel (b) shows results for an alternative model using AUC difference as the sole predictor. The single line shows predicted probability of choosing the target option by levels of AUC difference, with 95\% confidence band. The results show that as the AUC difference increases, probability of choosing the target option increases strongly. Together, the two panels confirm strong impacts of AUC difference on probability of choosing the target image.}
        \label{fig:analysisT1} 
\end{figure}

\para{Task 2} In \ref{Task:T2}, we predicted subjects would choose the higher-saliency scatterplot as showing clearer cluster structure. The probability of choosing this target option should increase as differences in threshold bar lengths between the two response options grow. As described above, we modeled that probability that the subject would choose the target (higher-saliency) option in trial $t$ using the bar lengths for the target (TLength) and comparison (lower-saliency; CLength) options and included random intercepts for each respondent ($i$) and stimulus dataset ($j$) to account for dependency across trials:

\begin{align}
ChooseTarget_{ijt} &\sim Bernoulli(p_{ijt}) \\
logit(p_{ijt}) &= \beta_{1} \times TLength_{ijt} + \beta_{2} \times CLength_{ijt}\ + \nonumber \\  
&\qquad \alpha + \gamma_{i} + \delta_{j} \nonumber \\
\gamma_{i} &\sim Normal(\mu_{\gamma}, \sigma_{\gamma}) \nonumber \\  
\delta_{j} &\sim Normal( \mu_{\delta}, \sigma_{\delta}) \nonumber 
\end{align}

\para{Analysis Software} We fit models using the \textit{glmmTMB} package~\cite[v. 1.1.1]{glmmTMBarticle} in \emph{R}~\cite[v. 4.1.0]{RbasePkg}. We computed and formatted model results using the \textit{modelbased}~\cite{modelbasedPkg} and \textit{parameters}~\cite{paramatersArticle, parametersPkg} packages. We managed data using the \textit{dplyr}~\cite{dplyrPkg} and \textit{readr}~\cite{readrPkg} packages. We visualized model results using the \textit{see}~\cite{seeArticle}, \textit{ggdist}, \textit{ggplot2}, \textit{ggtext}, and \textit{patchwork}~\cite{patchworkPkg} packages.

\subsection{Results}

\para{Task 1} Results for \ref{Task:T1} are shown in \autoref{tbl:modelpredictionsT1} and \autoref{fig:analysisT1}. 
\autoref{tbl:modelpredictionsT1} shows predicted probabilities of choosing the target option (Pr(CT)) based on the fitted model, with 95\% confidence intervals, for combinations of target and comparison AUC values. \autoref{fig:analysisT1} visualizes these results using a scatterplot showing individual trials with lines showing predicted probabilities of choosing the target option as a function of difference between the AUC values for the two scatterplot stimuli as options.

As shown, AUC differences strongly affected subject choice of which scatterplot was more similar to the reference plot. When the two stimuli options were equally SR (or DS) to the reference plot (i.e., when the AUC difference between the two plots' AUC values $\approx 0$), participants tended to select both the target option and the comparison option at similar rates (Pr(CT) = .61 [.43, .77]). However, when the target plot was much more similar to the reference plot than the comparison plot (e.g., when the difference between their AUC values $=4$), participants were much more likely to choose the target scatterplot (Pr(Ct) = .95 [.99, .85]) when AUC difference $=3$. This effect did not substantially vary across absolute levels of the target plot AUC (e.g., predicted probabilities for an AUC difference $=2$ were similar regardless of whether the target plot was highly similar or only somewhat similar to the reference plot); this indicates that it is the difference in AUC values between the options that drives the change in subject choices. \ref{Hyp:H1} is validated.

%
%
%
%
%

\begin{table}[!t] 
    \centering
    \caption{Model predictions for \ref{Task:T2} show predicted probabilities of choosing the target option (Pr(CT)) based on the fitted model, with 95\% CI, for combinations of target and comparison threshold bar lengths. Low indicates short bar length (low saliency), Med indicates medium bar length (medium saliency), High indicates long bar length (high saliency).}
    \label{tbl:modelpredictionsT2}
    
    \resizebox{0.95\linewidth}{!}{%
    \begin{tabular}{>{\raggedleft\arraybackslash}p{.09\textwidth}|>{\raggedleft\arraybackslash}p{.09\textwidth}|>{\centering\arraybackslash}p{.056\textwidth}|>{\centering\arraybackslash}p{.053\textwidth}|>{\centering\arraybackslash}p{.035\textwidth}|>{\centering\arraybackslash}p{.085\textwidth}}
\toprule
   \multicolumn{1}{c|}{Target} & \multicolumn{1}{c|}{Comp.} & Length & & & \\ 
   \multicolumn{1}{c|}{length} & \multicolumn{1}{c|}{length} & diff. & Pr(CT) & SE & 95\% CI \\
\hline
Low {[}0.15{]}{}{}{}{} & Low {[}0.15{]}{}{}{}{} & 0.00 & 0.66 & 0.06 & {[}0.54, 0.76{]} \\
Med.\ {[}0.50{]}{}{}{}{} & Med.\ {[}0.50{]}{}{}{}{} & 0.00 & 0.62 & 0.06 & {[}0.51, 0.73{]} \\
High {[}1.50{]}{}{}{}{} & High {[}1.50{]}{}{}{}{} & 0.00 & 0.52 & 0.07 & {[}0.38, 0.66{]} \\
Med.\ {[}0.50{]}{}{}{}{} & Low {[}0.15{]}{}{}{}{} & 0.35 & 0.68 & 0.05 & {[}0.56, 0.77{]} \\
High {[}1.50{]}{}{}{}{} & Med.\ {[}0.50{]}{}{}{}{} & 1.00 & 0.68 & 0.05 & {[}0.57, 0.78{]} \\
High {[}1.50{]}{}{}{}{} & Low {[}0.15{]}{}{}{}{} & 1.35 & 0.73 & 0.05 & {[}0.62, 0.82{]} \\
\bottomrule
\end{tabular}}

\vspace{3pt}
{\footnotesize Pr(CT): Pr(ChooseTarget)}
\end{table}

%
%
%
%
\begin{figure}[!t]
    \centering
      
    {\includegraphics[width=0.95\linewidth]{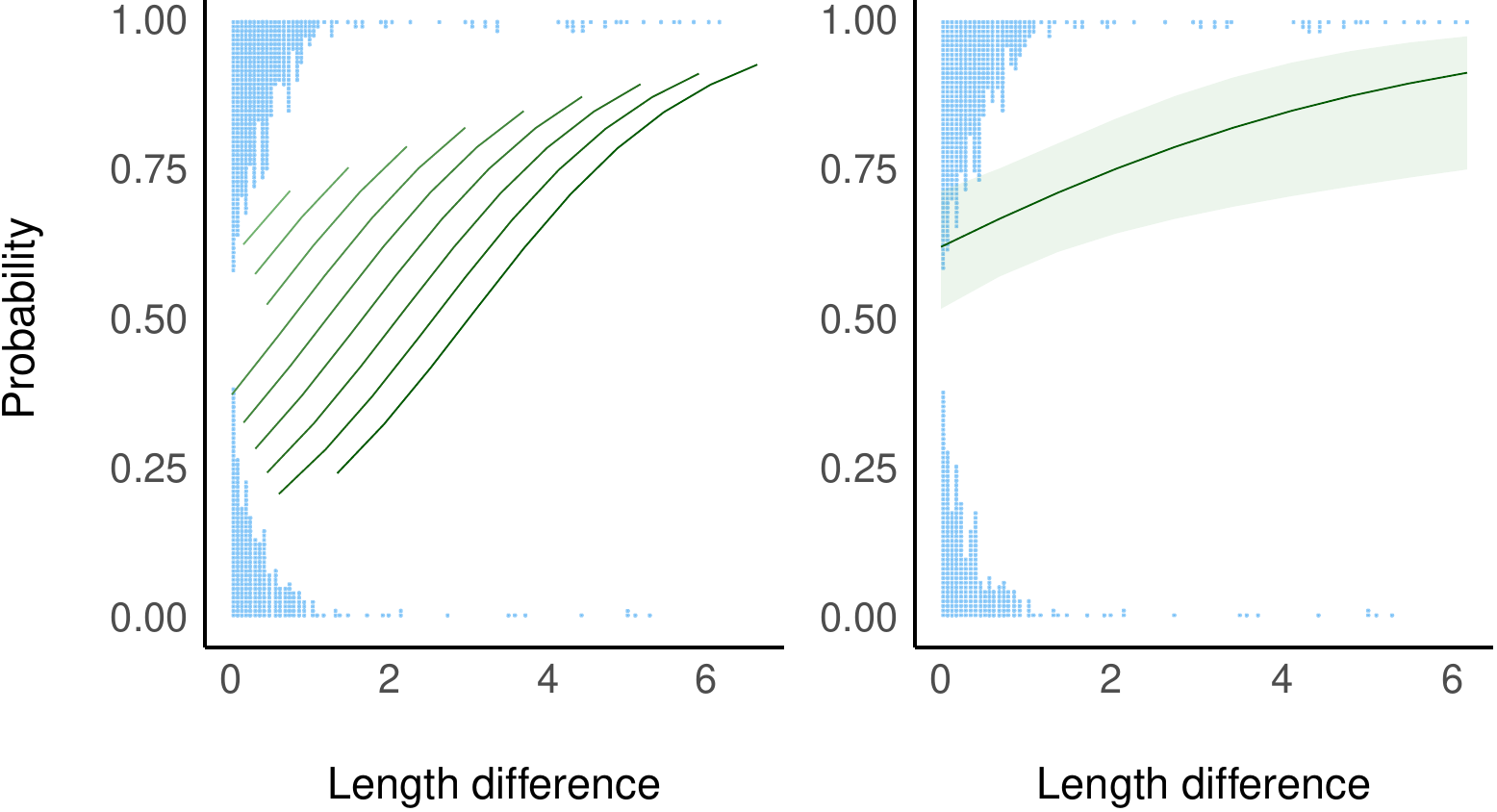}\hspace{0.025\linewidth}}

    \vspace{3pt}
    \hspace{10pt}
    \begin{minipage}[b]{.427\linewidth}
        \subcaption{Main model predictions: darker line indicates higher length for target plot.}
    \end{minipage}
    \hspace{5pt}
    \begin{minipage}[b]{.425\linewidth}
        \subcaption{Predictions and 95\% CI from model with length difference as the predictor.}
    \end{minipage}
    
    \caption{User study analysis for \ref{Task:T2}, Pr(ChooseTarget) by threshold bar length difference. The blue dot represents data points for each observation and lines show predicted probabilities of choosing the target option as a function of difference between the bar lengths for the two options.
    Panel~(a) shows results for the main model. Each line reflects a combination of Target length and Comp.\ length as shown in \autoref{tbl:modelpredictionsT2}. Darker lines indicate higher values of Target length. Each line shows that as the length difference increases, probability of choosing the target option increases strongly. However, the impact of length difference is most pronounced when the when the Target length is high (dark green lines), indicating that saliency contrast between images is most impactful when the target option has high saliency. Panel~(b) shows results for an alternative model using length difference as the sole predictor. The line shows predicted probability of choosing the target option by levels of length difference, with 95\% confidence band. This line shows that as the length difference increases, probability of choosing the target option increases. The slope of this line is somewhat shallow. Together, the two panels confirm strong impacts of length (saliency) difference on probability of choosing the target image, but also that such contrasts are most important when the target image has high saliency.}
    \label{fig:analysisT2} 
\end{figure}
%

\para{Task 2} Results for \ref{Task:T2} are shown in \autoref{tbl:modelpredictionsT2} and \autoref{fig:analysisT2}. \autoref{tbl:modelpredictionsT2} shows predicted probabilities of choosing the target option (Pr(CT)) based on the fitted model, with 95\% confidence intervals, for combinations of target and comparison threshold bar lengths. In the table, Low indicates short bar length (low saliency), Med indicates medium bar length (medium saliency), High indicates long bar length (high saliency). \autoref{fig:analysisT2} shows the results of the individual trials in a scatterplot with lines showing predicted probabilities of choosing the target option as a function of difference between the bar lengths for the two options.

Bar length strongly affected participants' choice of which scatterplot was clearer. Participants were much more likely to choose the target scatterplot as having clearer cluster structure when there was a large difference in bar lengths between the two scatterplots. Participants were much more likely to choose the target scatterplot as having clearer cluster structure when there was a large difference in threshold bar lengths between the two scatterplots (e.g., .73 [.62, .82] for a length difference of 1.35 versus .52 [.38, .66] for a length difference of $0$). Thus, saliency differences were a strong predictor of subject perceptions of cluster structure clarity. \ref{Hyp:H2} is validated. Bar length differences also strongly affected subject choice of which scatterplot showed clearer cluster structure.

\section{Case Study}
\label{sec-label:casestudy}

To evaluate the quality of results and utility of the interface, we conducted a case study. We recruited ten graduate students from our departments who are researching visualization or taken a data visualization course but had not been previously exposed to our project or interface. To compare the utility of our interface and perform qualitative analysis, we also constructed a \href{http://scatter.projects.jadorno.com/vis-2}{\textit{manual optimization interface}}, further referred to as M, which is different from our \textit{user-guided optimization interface}, further referred to as A (see \autoref{sec-label:model}). The M  interface featured sliders and option buttons for selecting scatterplot parameters, include sampling technique, number of points, point size, and opacity value.

We used the same datasets as in the user study (see \autoref{sec-label:userstudy}). Each participant was asked to use the M and A interface for 3 datasets each to design a scatterplot.  Dataset that was used for A and M were swapped between participants. The objective for each task (i.e., for each dataset) was to use the interface to select the factors (SA, SR, PS, and OP) that best highlight the clustering structure. The study was conducted in three parts: (1)~instruction and training, (2)~selecting optimal scatterplot, and (3)~interview. The total time for each participant was approximately one hour. Each participants assigned eight datasets: four for M (one for training and three for tasks) and four for A (one for training and three for tasks).

\begin{figure}[!ht]
    \centering
    \hspace{15pt}\includegraphics[width=0.4\linewidth]{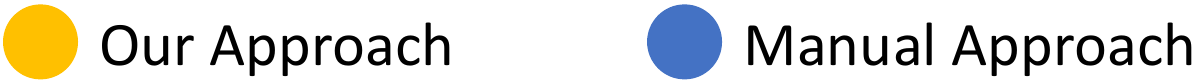}
    
    \vspace{5pt}
    \begin{minipage}[b]{.427\linewidth}
        \includegraphics[width=\linewidth]{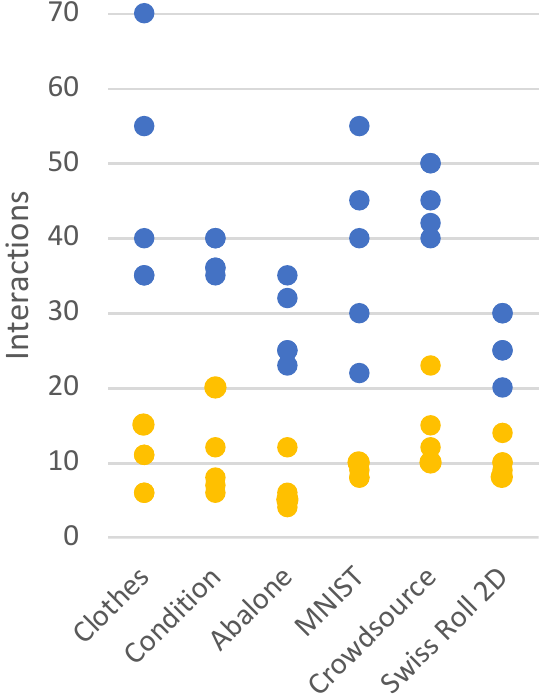}
        \subcaption{Count of Interactions}
        \label{fig:casestudy:interaction}
    \end{minipage}
    \hspace{10pt}
    \begin{minipage}[b]{.425\linewidth}
        \includegraphics[width=\linewidth]{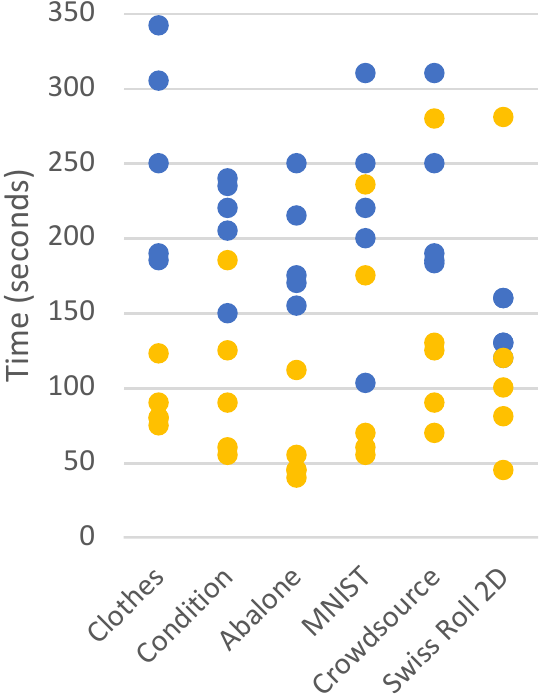}
        \subcaption{Time Taken}
        \label{fig:casestudy:time}
    \end{minipage}
    
    \caption{Case study participants' performance in terms of the number of interactions and time for six different datasets. Each dot represents one participant's results (some are obscured by overlap). Each column represents one dataset.}
    \label{fig:casestudy}
\end{figure}

\begin{figure*}[!t]
    \includegraphics[width=0.975\linewidth]{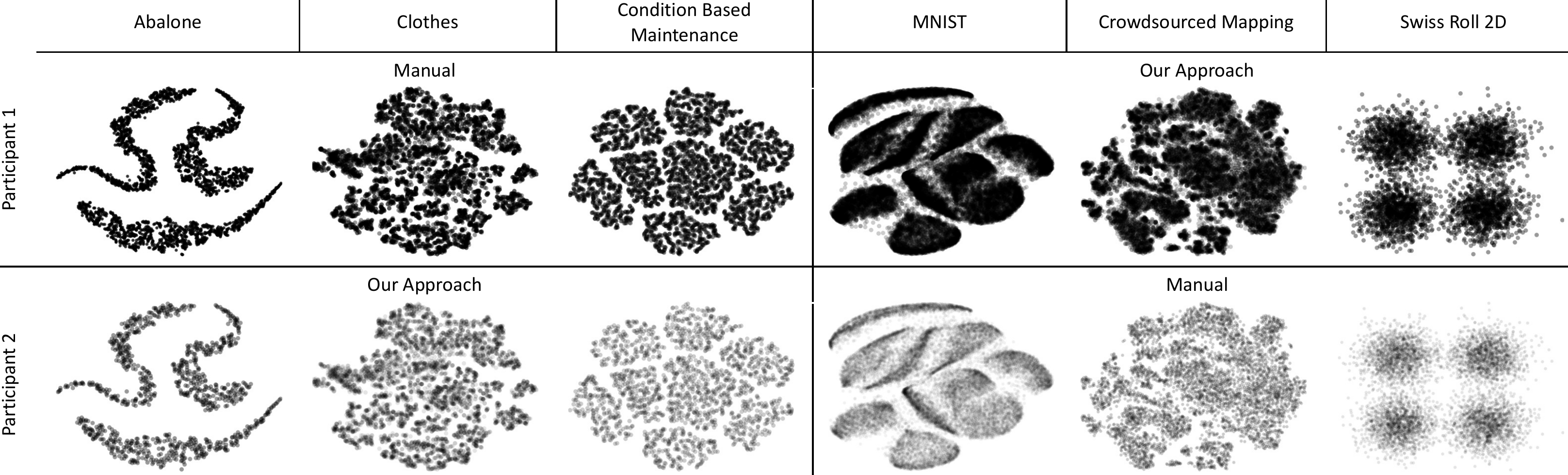}
    \caption{Images created by two participants in the case study. Each image represents the optimized design chosen by the participants using a manual or our automatic approach.}
    \label{fig:casestudy_images}
\end{figure*}

The results of the study can be seen in \autoref{fig:casestudy}, and the scatterplots from several subjects are shown in \autoref{fig:casestudy_images}. We first investigate the number of interactions required, where an interaction is defined as selecting the values of the factors to select an optimal scatterplot. As one can observe in \autoref{fig:casestudy:interaction}, the manual optimization required a significantly higher number of interactions. In addition, in terms of time, we saw that the manual approach also tended to require more time from participants (see \autoref{fig:casestudy:time}). From conversations with participants, we hypothesize time is related to their confidence (less time, higher confidence) in the optimality of their choice, whereas the number of interactions is related to the usability of the interface (fewer interactions, higher usability).

In terms of quality, since all the participants had different datasets for the manual and automatic methods, we could not compare between subjects. However, \autoref{fig:casestudy_images} shows the output images for six datasets from two of the participants. Without the labeling (see \autoref{fig:casestudy_images}), it is difficult to distinguish which are found using our interface (A) and which are found using the manual (M) approach. Second, each participant seemed to have their own preferred aesthetic, which they were able to produce in each interface.

\section{Discussion}
\label{sec-label:discussion}

The goal of our approach is to suggest an optimized visualization design to improve the effectiveness of the task performance, and it is important to understand how designers can use our models to reduce ambiguity in the data and thereby reduce the chance of misinterpretation, e.g., by having a visualization that is too sparse or over-saturated. Our approach uses a data-driven framework to compare and observe how cluster patterns change with a variety of density-influencing parameters of scatterplots, including point size and opacity visual encodings, as well as the subsampling algorithm and sampling rate. Our approach provides scatterplots ranked in order of their cluster saliency, where the saliency score (longest bar in the threshold plot) is a proxy for the clarity of the cluster structure in the scatterplot.

\subsection{Saliency as a Proxy of Cluster Structure} 
The theoretical models of Sadahiro state that proximity, and number and concentration of points, and density change affect cluster perception~\cite{sadahiro1997cluster}. Other experimental work has shown that the choice of visual factors which influence the visual density of scatterplot can have a significant effect on cluster identification~\cite{quadri2020modeling}. The threshold plot is computed on the visual density estimate of the scatterplots and identifies how clusters visually merge together, fitting well with the known factors that influence clustering. With each bar of the threshold plot we measure the saliency of that number of clusters. In other words, how likely it is that a user will see that number of clusters. Therefore, by identifying the longest bar we capture the cluster structure most likely visible to the user.

\subsection{Ambiguous Clustering Structure}  
Every dataset has an inherent properties that influence the visualization of the data. For example, data distribution plays a vital role in point concentrations concerning over-saturation or sparse distribution. Such properties often influence the visualization, leading to an ambiguous conclusion for even the optimal design choice (e.g., for Clothes and Crowdsourced Mapping datasets in \autoref{fig:casestudy_images}). For these data, optimizing the design has negligible effects on clarifying the cluster structure which remains ambiguous for most parameter configurations. This problem also has a weak relationship with bin size at the time of threshold plot generation. Bins that are too large smooth the result, while bins too small create noise. This problem has been partially investigated previously~\cite{quadri2020modeling} but deserves more attention in the future.

\subsection{Comparisons to Existing Approach}
\textit{Visual Quality Measures (VQMs)} are ideally based on perceptual models rather than heuristics and computational approaches. The existing approaches, such as \cite{abbas2019clustme, aupetit2019toward, sedlmair2015data}, apply measures that imitate how humans would score views (e.g., one or more clusters but not the specific count) based on the perceived patterns and can be used to accurately predict perceptual judgments. ClustMe used VQMs to model human judgments to rank scatterplots~\cite{abbas2019clustme} which was further extended in~\cite{aupetit2019toward}. These studies performed well in reproducing human judgments for  for quantifying cluster patterns as per points positions, but ignored the visual aspects (marker size, opacity, and visual density). Similarly, the \textit{scagnostics} technique utilizes density property that identifies concentrations of points, which is directly influenced by the distribution of points~\cite{wilkinson2005graph} to investigate the patterns. 

In contrast to these approach, our work proposes saliency score as a VQM that can be used to optimize design factors (data aspects and the visual encoding) and ranks the scatterplot designs on the cluster count that matches human understanding. It is important to note that the optimal design seemed to be both quantitative and qualitative. In other words, our saliency measure provided visualizations with clearer clustering structure, but each participant in our case study also seemed to have their own preferred aesthetic, which they were still able to produce with our interface.

\subsection{Limitations} 
Our approach has some limitations. First, we have not considered some other factors that could influence performance in either model, e.g., chart size, screen resolutions, etc. We have also not extensively analyzed time variance between individuals' performance on the datasets and their sampling rate. We have not explored the histogram bin size to compute the density model, but the same is extensively discussed in \cite{quadri2020modeling}. A final limitation is that we have not considered the relationship of our approach to confidence \cite{sacha2015role}, which is highly related to the nature of data~\cite{etemadpour2014perception}.

\section{Conclusions}
 
Scatterplots are among the most powerful and most widely used techniques for visual data exploration of 2D data. Design choices in visualization, scatterplots in this case, such as the visual encodings or data aspects, can directly impact the quality of decision-making for low-level tasks, such as clustering. 
 
We propose here a user-guided tool to optimize the design factors of scatterplot for salient cluster structure. By constructing frameworks, such as this one, that consider both the perceptions of the visual encodings and the task being performed enables maximizing the efficacy of the visualization. Our interactive tool leverages the application of the merge tree data structure to optimize the design decisions on sampling algorithms, sampling rate, symbol size, and opacity. We further validate our results with a user study, case studies, and demo interface that demonstrate guidelines that practitioners and designers can extend to other tasks on scatterplots.

\vspace{4pt}
\noindent  
Interface: {\small\textless\textcolor{blue}{\url{http://scatter.projects.jadorno.com/}\textgreater}}\\
Data: {\small\textless\textcolor{blue}{\url{https://osf.io/cxgq2/}\textgreater}}



\ifCLASSOPTIONcompsoc
  \section*{Acknowledgments}
\else
  \section*{Acknowledgment}
\fi

The project is supported in part by the National Science Foundation IIS-1845204 and National Science Foundation CNS-2127309 to the Computing Research Association for the CIFellows program.

\ifCLASSOPTIONcaptionsoff
  \newpage
\fi



\bibliographystyle{IEEEtran}
\bibliography{refs.bib}

\begin{thebibliography}{100}
\providecommand{\url}[1]{#1}
\csname url@samestyle\endcsname
\providecommand{\newblock}{\relax}
\providecommand{\bibinfo}[2]{#2}
\providecommand{\BIBentrySTDinterwordspacing}{\spaceskip=0pt\relax}
\providecommand{\BIBentryALTinterwordstretchfactor}{4}
\providecommand{\BIBentryALTinterwordspacing}{\spaceskip=\fontdimen2\font plus
\BIBentryALTinterwordstretchfactor\fontdimen3\font minus
  \fontdimen4\font\relax}
\providecommand{\BIBforeignlanguage}[2]{{%
\expandafter\ifx\csname l@#1\endcsname\relax
\typeout{** WARNING: IEEEtran.bst: No hyphenation pattern has been}%
\typeout{** loaded for the language `#1'. Using the pattern for}%
\typeout{** the default language instead.}%
\else
\language=\csname l@#1\endcsname
\fi
#2}}
\providecommand{\BIBdecl}{\relax}
\BIBdecl

\bibitem{friendly2005early}
M.~Friendly and D.~Denis, ``The early origins and development of the
  scatterplot,'' \emph{J.\ of the History of the Behavioral Sciences}, 2005.

\bibitem{quadri2021survey}
G.~J. Quadri and P.~Rosen, ``A survey of perception-based visualization studies
  by task,'' \emph{IEEE Transactions on Visualization and Computer Graphics},
  2021.

\bibitem{nguyen2016visual}
H.~Nguyen, P.~Rosen, and B.~Wang, ``Visual exploration of multiway dependencies
  in multivariate data,'' in \emph{SIGGRAPH ASIA Symp.\ on Visualization},
  2016.

\bibitem{rensink2010perception}
R.~Rensink and G.~Baldridge, ``The perception of correlation in scatterplots,''
  \emph{Computer Graphics Forum}, 2010.

\bibitem{harrison2014ranking}
L.~Harrison, F.~Yang, S.~Franconeri, and R.~Chang, ``Ranking visualizations of
  correlation using weber's law.'' \emph{IEEE Trans.\ on Visualization and
  Comp.\ Graphics}, 2014.

\bibitem{gleicher2013perception}
M.~Gleicher, M.~Correll, C.~Nothelfer, and S.~Franconeri, ``Perception of
  average value in multiclass scatterplots,'' \emph{IEEE Trans.\ on
  Visualization and Comp.\ Graphics}, 2013.

\bibitem{sarikaya2018design}
A.~Sarikaya, M.~Gleicher, and D.~Szafir, ``Design factors for summary
  visualization in visual analytics,'' \emph{Computer Graphics Forum}, 2018.

\bibitem{quadri2020modeling}
G.~J. Quadri and P.~Rosen, ``Modeling the influence of visual density on
  cluster perception in scatterplots using topology,'' \emph{IEEE Trans.\ on
  Visualization and Comp.\ Graphics}, 2020.

\bibitem{munzner2014visualization}
T.~Munzner, \emph{Visualization Analysis and Design}.\hskip 1em plus 0.5em
  minus 0.4em\relax CRC press, 2014.

\bibitem{amar2005low}
R.~Amar, J.~Eagan, and J.~Stasko, ``Low-level components of analytic activity
  in information visualization,'' in \emph{IEEE Symp.\ on Information
  Visualization (InfoVis)}, 2005.

\bibitem{kim2018assessing}
Y.~Kim and J.~Heer, ``Assessing effects of task and data distribution on the
  effectiveness of visual encodings,'' \emph{Computer Graphics Forum}, 2018.

\bibitem{micallef2017towards}
L.~Micallef, G.~Palmas, A.~Oulasvirta, and T.~Weinkauf, ``Towards perceptual
  optimization of the visual design of scatterplots,'' \emph{IEEE Trans.\ on
  Visualization and Comp.\ Graphics}, 2017.

\bibitem{szafir2018modeling}
D.~Szafir, ``Modeling color difference for visualization design,'' \emph{IEEE
  Trans.\ on Visualization and Comp.\ Graphics}, 2018.

\bibitem{sedlmair2012taxonomy}
M.~Sedlmair, A.~Tatu, T.~Munzner, and M.~Tory, ``A taxonomy of visual cluster
  separation factors,'' \emph{Computer Graphics Forum}, 2012.

\bibitem{ellis2007taxonomy}
G.~Ellis and A.~Dix, ``A taxonomy of clutter reduction for information
  visualisation,'' \emph{IEEE Trans.\ on Visualization and Comp.\ Graphics},
  2007.

\bibitem{sarikaya2018scatterplots}
A.~Sarikaya and M.~Gleicher, ``Scatterplots: Tasks, data, and designs,''
  \emph{IEEE Trans.\ on Visualization and Comp.\ Graphics}, 2018.

\bibitem{aupetit2019toward}
M.~Aupetit, M.~Sedlmair, M.~M. Abbas, A.~Baggag, and H.~Bensmail, ``Toward
  perception-based evaluation of clustering techniques for visual analytics,''
  in \emph{2019 IEEE Visualization Conference (VIS)}.\hskip 1em plus 0.5em
  minus 0.4em\relax IEEE, 2019, pp. 141--145.

\bibitem{matute2017skeleton}
J.~Matute, A.~Telea, and L.~Linsen, ``Skeleton-based scagnostics,'' \emph{IEEE
  Trans.\ on Visualization and Comp.\ Graphics}, 2017.

\bibitem{sedlmair2015data}
M.~Sedlmair and M.~Aupetit, ``Data-driven evaluation of visual quality
  measures,'' in \emph{Computer Graphics Forum}, vol.~34, no.~3.\hskip 1em plus
  0.5em minus 0.4em\relax Wiley Online Library, 2015, pp. 201--210.

\bibitem{ma2018scatternet}
Y.~Ma, A.~Tung, W.~Wang, X.~Gao, Z.~Pan, and W.~Chen, ``Scatternet: A deep
  subjective similarity model for visual analysis of scatterplots,'' \emph{IEEE
  Trans.\ on Visualization and Comp.\ Graphics}, 2018.

\bibitem{dang2014transforming}
T.~Dang and L.~Wilkinson, ``Transforming scagnostics to reveal hidden
  features,'' \emph{IEEE Trans.\ on Visualization and Comp.\ Graphics}, 2014.

\bibitem{pandey2016towards}
A.~Pandey, J.~Krause, C.~Felix, J.~Boy, and E.~Bertini, ``Towards understanding
  human similarity perception in the analysis of large sets of scatter plots,''
  in \emph{ACM SIGCHI Conference on Human Factors in Computing}, 2016.

\bibitem{rensink2014prospects}
R.~Rensink, ``On the prospects for a science of visualization,'' in
  \emph{Handbook of human centric visualization}.\hskip 1em plus 0.5em minus
  0.4em\relax Springer, 2014.

\bibitem{wang2018optimizing}
Y.~Wang, X.~Chen, T.~Ge, C.~Bao, M.~Sedlmair, C.-W. Fu, O.~Deussen, and
  B.~Chen, ``Optimizing color assignment for perception of class separability
  in multiclass scatterplots,'' \emph{IEEE Trans.\ on Visualization and Comp.\
  Graphics}, 2018.

\bibitem{lu2020palettailor}
K.~Lu, M.~Feng, X.~Chen, M.~Sedlmair, O.~Deussen, D.~Lischinski, Z.~Cheng, and
  Y.~Wang, ``Palettailor: discriminable colorization for categorical data,''
  \emph{IEEE Transactions on Visualization and Computer Graphics}, vol.~27,
  no.~2, pp. 475--484, 2020.

\bibitem{wang2018line}
Y.~Wang, F.~Han, L.~Zhu, O.~Deussen, and B.~Chen, ``Line graph or scatter plot?
  automatic selection of methods for visualizing trends in time series,''
  \emph{IEEE Trans.\ on Visualization and Comp.\ Graphics}, 2018.

\bibitem{abbas2019clustme}
M.~Abbas, M.~Aupetit, M.~Sedlmair, and H.~Bensmail, ``Clustme: A visual quality
  measure for ranking monochrome scatterplots based on cluster patterns,''
  \emph{Computer Graphics Forum}, 2019.

\bibitem{bederson2002ordered}
B.~Bederson, B.~Shneiderman, and M.~Wattenberg, ``Ordered and quantum treemaps:
  Making effective use of 2d space to display hierarchies,'' \emph{ACM
  Transactions on Graphics}, 2002.

\bibitem{derthick2003constant}
M.~Derthick, M.~Christel, A.~Hauptmann, and H.~Wactlar, ``Constant density
  displays using diversity sampling,'' in \emph{IEEE Symp.\ on Information
  Visualization (InfoVis)}, 2003.

\bibitem{plaisant1996lifelines}
C.~Plaisant, B.~Milash, A.~Rose, S.~Widoff, and B.~Shneiderman, ``Lifelines:
  visualizing personal histories,'' in \emph{ACM SIGCHI Conference on Human
  Factors in Computing}, 1996.

\bibitem{woodruff1998constant}
A.~Woodruff, J.~Landay, and M.~Stonebraker, ``Constant density visualizations
  of non-uniform distributions of data,'' in \emph{ACM Symp.\ on User interface
  software and technology}, 1998.

\bibitem{matejka2015dynamic}
J.~Matejka, F.~Anderson, and G.~Fitzmaurice, ``Dynamic opacity optimization for
  scatter plots,'' in \emph{ACM Conference on Human Factors in Computing
  Systems}, 2015.

\bibitem{wegman1997high}
E.~Wegman and Q.~Luo, ``High dimensional clustering using parallel coordinates
  and the grand tour,'' in \emph{Classification and Knowledge
  Organization}.\hskip 1em plus 0.5em minus 0.4em\relax Springer, 1997.

\bibitem{kosara2002focus+}
R.~Kosara, S.~Miksch, and H.~Hauser, ``Focus+ context taken literally,''
  \emph{IEEE Computer Graphics and Applications}, 2002.

\bibitem{johansson2006revealing}
J.~Johansson, P.~Ljung, M.~Jern, and M.~Cooper, ``Revealing structure in
  visualizations of dense 2d and 3d parallel coordinates,'' \emph{Information
  Visualization}, 2006.

\bibitem{fekete2002interactive}
J.-D. Fekete and C.~Plaisant, ``Interactive information visualization of a
  million items,'' in \emph{IEEE Symp.\ on Information Visualization
  (InfoVis)}, 2002.

\bibitem{cohen1973eta}
J.~Cohen, ``Eta-squared and partial eta-squared in communication science,''
  \emph{Human Communication Research}, 1973.

\bibitem{chen2014visual}
H.~Chen, W.~Chen, H.~Mei, Z.~Liu, K.~Zhou, W.~Chen, W.~Gu, and K.-L. Ma,
  ``Visual abstraction and exploration of multi-class scatterplots,''
  \emph{IEEE Trans.\ on Visualization and Comp.\ Graphics}, 2014.

\bibitem{urribarri2017prediction}
D.~Urribarri and S.~Castro, ``Prediction of data visibility in two-dimensional
  scatterplots,'' \emph{Information Visualization}, 2017.

\bibitem{hu2019data}
R.~Hu, T.~Sha, O.~Van~Kaick, O.~Deussen, and H.~Huang, ``Data sampling in
  multi-view and multi-class scatterplots via set cover optimization,''
  \emph{IEEE Trans.\ on Visualization and Comp.\ Graphics}, 2019.

\bibitem{szafir2016four}
D.~Szafir, S.~Haroz, M.~Gleicher, and S.~Franconeri, ``Four types of ensemble
  coding in data visualizations,'' \emph{J.\ of Vision}, 2016.

\bibitem{sadahiro1997cluster}
Y.~Sadahiro, ``Cluster perception in the distribution of point objects,''
  \emph{Cartographica: The International Journal for Geographic Information and
  Geovisualization}, 1997.

\bibitem{few2008solutions}
S.~Few and P.~Edge, ``Solutions to the problem of over-plotting in graphs,''
  \emph{Visual Business Intelligence Newsletter}, 2008.

\bibitem{correll2018looks}
M.~Correll, M.~Li, G.~Kindlmann, and C.~Scheidegger, ``Looks good to me:
  Visualizations as sanity checks,'' \emph{IEEE Trans.\ on Visualization and
  Comp.\ Graphics}, 2018.

\bibitem{ellis2002density}
G.~Ellis and A.~Dix, ``Density control through random sampling: an
  architectural perspective,'' in \emph{International Conference on Information
  Visualisation (IV)}, 2002.

\bibitem{dix2002chance}
A.~Dix and G.~Ellis, ``By chance enhancing interaction with large data sets
  through statistical sampling,'' in \emph{Working Conference on Advanced
  Visual Interfaces}, 2002.

\bibitem{bertini2005improving}
E.~Bertini and G.~Santucci, ``Improving 2d scatterplots effectiveness through
  sampling, displacement, and user perception,'' in \emph{International
  Conference on Information Visualisation (IV)}, 2005.

\bibitem{yuan2020evaluation}
J.~Yuan, S.~Xiang, J.~Xia, L.~Yu, and S.~Liu, ``Evaluation of sampling methods
  for scatterplots,'' \emph{IEEE Trans.\ on Visualization and Comp.\ Graphics},
  2020.

\bibitem{carr1987scatterplot}
D.~Carr, R.~Littlefield, W.~Nicholson, and J.~Littlefield, ``Scatterplot matrix
  techniques for large n,'' \emph{J.\ of the American Statistical Association},
  1987.

\bibitem{bachthaler2008continuous}
S.~Bachthaler and D.~Weiskopf, ``Continuous scatterplots,'' \emph{IEEE Trans.\
  on Visualization and Comp.\ Graphics}, 2008.

\bibitem{keim2010generalized}
D.~Keim, M.~Hao, U.~Dayal, H.~Janetzko, and P.~Bak, ``Generalized scatter
  plots,'' \emph{Information Visualization}, 2010.

\bibitem{mayorga2013splatterplots}
A.~Mayorga and M.~Gleicher, ``Splatterplots: Overcoming overdraw in scatter
  plots,'' \emph{IEEE Trans.\ on Visualization and Comp.\ Graphics}, 2013.

\bibitem{trautner2020sunspot}
T.~Trautner, F.~Bolte, S.~Stoppel, and S.~Bruckner, ``Sunspot plots:
  Model-based structure enhancement for dense scatter plots,'' \emph{Computer
  Graphics Forum}, 2020.

\bibitem{rojas2017sampling}
J.~Rojas, M.~Kery, S.~Rosenthal, and A.~Dey, ``Sampling techniques to improve
  big data exploration,'' in \emph{IEEE Symp.\ on Large Data Analysis and
  Visualization (LDAV)}, 2017.

\bibitem{palmer2000density}
C.~Palmer and C.~Faloutsos, ``Density biased sampling: An improved method for
  data mining and clustering,'' in \emph{ACM SIGMOD International Conference on
  Management of Data}, 2000.

\bibitem{bertini2006give}
E.~Bertini and G.~Santucci, ``Give chance a chance: modeling density to enhance
  scatter plot quality through random data sampling,'' \emph{Information
  Visualization}, 2006.

\bibitem{bertini2004chance}
------, ``By chance is not enough: preserving relative density through
  nonuniform sampling,'' in \emph{Conference on Information Visualisation},
  2004.

\bibitem{joia2015uncovering}
P.~Joia, F.~Petronetto, and L.~G. Nonato, ``Uncovering representative groups in
  multidimensional projections,'' \emph{Computer Graphics Forum}, 2015.

\bibitem{zheng2013quality}
Y.~Zheng, J.~Jestes, J.~Phillips, and F.~Li, ``Quality and efficiency for
  kernel density estimates in large data,'' in \emph{ACM SIGMOD International
  Conference on Management of Data}, 2013.

\bibitem{chen2019recursive}
X.~Chen, T.~Ge, J.~Zhang, B.~Chen, C.-W. Fu, O.~Deussen, and Y.~Wang, ``A
  recursive subdivision technique for sampling multi-class scatterplots,''
  \emph{IEEE Trans.\ on Visualization and Comp.\ Graphics}, 2019.

\bibitem{xia2017ldsscanner}
J.~Xia, F.~Ye, W.~Chen, Y.~Wang, W.~Chen, Y.~Ma, and A.~Tung, ``Ldsscanner:
  Exploratory analysis of low-dimensional structures in high-dimensional
  datasets,'' \emph{IEEE Trans.\ on Visualization and Comp.\ Graphics}, 2017.

\bibitem{dos2012ilamp}
E.~dos Santos~Amorim, E.~Brazil, J.~Daniels, P.~Joia, L.~Nonato, and M.~Sousa,
  ``ilamp: Exploring high-dimensional spacing through backward multidimensional
  projection,'' in \emph{IEEE Visual Analytics Science and Technology (VAST)},
  2012.

\bibitem{poco2011framework}
J.~Poco, R.~Etemadpour, F.~V. Paulovich, T.~Long, P.~Rosenthal, M.~d. Oliveira,
  L.~Linsen, and R.~Minghim, ``A framework for exploring multidimensional data
  with 3d projections,'' \emph{Computer Graphics Forum}, 2011.

\bibitem{rieck2015persistent}
B.~Rieck and H.~Leitte, ``Persistent homology for the evaluation of
  dimensionality reduction schemes,'' \emph{Computer Graphics Forum}, 2015.

\bibitem{xiang2019interactive}
S.~Xiang, X.~Ye, J.~Xia, J.~Wu, Y.~Chen, and S.~Liu, ``Interactive correction
  of mislabeled training data,'' in \emph{IEEE Visual Analytics Science and
  Technology (VAST)}, 2019.

\bibitem{liu2017visual}
S.~Liu, J.~Xiao, J.~Liu, X.~Wang, J.~Wu, and J.~Zhu, ``Visual diagnosis of tree
  boosting methods,'' \emph{IEEE Trans.\ on Visualization and Comp.\ Graphics},
  2017.

\bibitem{zhao2019oui}
X.~Zhao, W.~Cui, Y.~Wu, H.~Zhang, H.~Qu, and D.~Zhang, ``Oui! outlier
  interpretation on multi-dimensional data via visual analytics,''
  \emph{Computer Graphics Forum}, 2019.

\bibitem{cheng2018colormap}
S.~Cheng, W.~Xu, and K.~Mueller, ``Colormap nd: A data-driven approach and tool
  for mapping multivariate data to color,'' \emph{IEEE Trans.\ on Visualization
  and Comp.\ Graphics}, 2018.

\bibitem{wei2008parallel}
L.-Y. Wei, ``Parallel poisson disk sampling,'' \emph{ACM Transactions on
  Graphics}, 2008.

\bibitem{wei2010multi}
------, ``Multi-class blue noise sampling,'' \emph{ACM Transactions on
  Graphics}, 2010.

\bibitem{berger2016cite2vec}
M.~Berger, K.~McDonough, and L.~Seversky, ``cite2vec: Citation-driven document
  exploration via word embeddings,'' \emph{IEEE Trans.\ on Visualization and
  Comp.\ Graphics}, 2016.

\bibitem{breunig2000lof}
M.~Breunig, H.-P. Kriegel, R.~Ng, and J.~Sander, ``Lof: identifying
  density-based local outliers,'' in \emph{ACM SIGMOD International Conference
  on Management of Data}, 2000.

\bibitem{yellott1983spectral}
J.~Yellott, ``Spectral consequences of photoreceptor sampling in the rhesus
  retina,'' \emph{Science}, 1983.

\bibitem{yan2015survey}
D.-M. Yan, J.-W. Guo, B.~Wang, X.-P. Zhang, and P.~Wonka, ``A survey of
  blue-noise sampling and its applications,'' \emph{J.\ of Computer Science and
  Technology}, 2015.

\bibitem{gramazio2014relation}
C.~Gramazio, K.~Schloss, and D.~Laidlaw, ``The relation between visualization
  size, grouping, and user performance,'' \emph{IEEE Trans.\ on Visualization
  and Comp.\ Graphics}, 2014.

\bibitem{cleveland1984graphical}
W.~Cleveland and R.~McGill, ``Graphical perception: Theory, experimentation,
  and application to the development of graphical methods,'' \emph{J.\ of the
  American Statistical Association}, 1984.

\bibitem{wong2010points}
B.~Wong, ``Points of view: gestalt principles,'' \emph{Nature Methods}, 2010.

\bibitem{cohen2008perceptual}
E.~H. Cohen, M.~Singh, and L.~Maloney, ``Perceptual segmentation and the
  perceived orientation of dot clusters: The role of robust statistics,''
  \emph{J.\ of Vision}, 2008.

\bibitem{anobile2016number}
G.~Anobile, G.~Cicchini, and D.~Burr, ``Number as a primary perceptual
  attribute: A review,'' \emph{Perception}, 2016.

\bibitem{wilkinson2005graph}
L.~Wilkinson, A.~Anand, and R.~Grossman, ``Graph-theoretic scagnostics,'' in
  \emph{IEEE Symp.\ on Information Visualization (InfoVis)}, 2005.

\bibitem{etemadpour2017density}
R.~Etemadpour and A.~G. Forbes, ``Density-based motion,'' \emph{Information
  Visualization}, 2017.

\bibitem{veras2019saliency}
R.~Veras and C.~Collins, ``Saliency deficit and motion outlier detection in
  animated scatterplots,'' in \emph{ACM SIGCHI Conference on Human Factors in
  Computing}, 2019.

\bibitem{chen2018using}
H.~Chen, S.~Engle, A.~Joshi, E.~D. Ragan, B.~Yuksel, and L.~Harrison, ``Using
  animation to alleviate overdraw in multiclass scatterplot matrices,'' in
  \emph{ACM SIGCHI Conference on Human Factors in Computing}, 2018.

\bibitem{heer2009sizing}
J.~Heer, N.~Kong, and M.~Agrawala, ``Sizing the horizon: the effects of chart
  size and layering on the graphical perception of time series
  visualizations,'' in \emph{ACM SIGCHI Conference on Human Factors in
  Computing}, 2009.

\bibitem{chung2016ordered}
D.~Chung, D.~Archambault, R.~Borgo, D.~Edwards, R.~Laramee, and M.~Chen, ``How
  ordered is it? on the perceptual orderability of visual channels,''
  \emph{Computer Graphics Forum}, 2016.

\bibitem{rosen2018hybrid}
P.~Rosen, J.~Tu, and L.~A. Piegl, ``A hybrid solution to parallel calculation
  of augmented join trees of scalar fields in any dimension,''
  \emph{Computer-Aided Design and Applications}, 2018.

\bibitem{zomorodian2005computing}
A.~Zomorodian and G.~Carlsson, ``Computing persistent homology,''
  \emph{Discrete \& Computational Geometry}, vol.~33, no.~2, pp. 249--274,
  2005.

\bibitem{lecun1998gradient}
Y.~LeCun, L.~Bottou, Y.~Bengio, and P.~Haffner, ``Gradient-based learning
  applied to document recognition,'' \emph{Proceedings of the IEEE}, 1998.

\bibitem{coraddu2016machine}
A.~Coraddu, L.~Oneto, A.~Ghio, S.~Savio, D.~Anguita, and M.~Figari, ``Machine
  learning approaches for improving condition-based maintenance of naval
  propulsion plants,'' \emph{Proceedings of the Institution of Mechanical
  Engineers, Part M: J.\ of Engineering for the Maritime Environment}, 2016.

\bibitem{johnson2016integrating}
B.~Johnson and K.~Iizuka, ``Integrating openstreetmap crowdsourced data and
  landsat time-series imagery for rapid land use/land cover (lulc) mapping:
  Case study of the laguna de bay area of the philippines,'' \emph{Applied
  Geography}, 2016.

\bibitem{andrzejak2001indications}
R.~Andrzejak, K.~Lehnertz, F.~Mormann, C.~Rieke, P.~David, and C.~Elger,
  ``Indications of nonlinear deterministic and finite-dimensional structures in
  time series of brain electrical activity: Dependence on recording region and
  brain state,'' \emph{Physical Review E}, 2001.

\bibitem{sedlmair2013empirical}
M.~Sedlmair, T.~Munzner, and M.~Tory, ``Empirical guidance on scatterplot and
  dimension reduction technique choices,'' \emph{IEEE Trans.\ on Visualization
  and Comp.\ Graphics}, 2013.

\bibitem{dua2017uci}
\BIBentryALTinterwordspacing
D.~Dua and C.~Graff, ``{UCI} machine learning repository,'' 2017. [Online].
  Available: \url{http://archive.ics.uci.edu/ml}
\BIBentrySTDinterwordspacing

\bibitem{kohavi1996scaling}
R.~Kohavi \emph{et~al.}, ``Scaling up the accuracy of naive-bayes classifiers:
  A decision-tree hybrid.'' in \emph{Kdd}, vol.~96, 1996, pp. 202--207.

\bibitem{kaya2019predicting}
H.~Kaya, P.~T{\"u}fekci, and E.~Uzun, ``Predicting co and no x emissions from
  gas turbines: novel data and a benchmark pems,'' \emph{Turkish journal of
  electrical engineering \& computer sciences}, vol.~27, no.~6, pp. 4783--4796,
  2019.

\bibitem{yeh2009comparisons}
I.-C. Yeh and C.-h. Lien, ``The comparisons of data mining techniques for the
  predictive accuracy of probability of default of credit card clients,''
  \emph{Expert systems with applications}, vol.~36, no.~2, pp. 2473--2480,
  2009.

\bibitem{strack2014impact}
B.~Strack, J.~P. DeShazo, C.~Gennings, J.~L. Olmo, S.~Ventura, K.~J. Cios, and
  J.~N. Clore, ``Impact of hba1c measurement on hospital readmission rates:
  analysis of 70,000 clinical database patient records,'' \emph{BioMed Research
  International}, 2014.

\bibitem{StatCounter2019}
StatCounter, ``Desktop screen resolution stats worldwide,''
  \url{http://gs.statcounter.com/screen-resolution-stats/desktop/worldwide},
  2019.

\bibitem{akcora2019bitcoinheist}
C.~G. Akcora, Y.~Li, Y.~R. Gel, and M.~Kantarcioglu, ``Bitcoinheist:
  Topological data analysis for ransomware detection on the bitcoin
  blockchain,'' \emph{arXiv preprint}, 2019.

\bibitem{glmmTMBarticle}
M.~Brooks, K.~Kristensen, K.~Van~Benthem, A.~Magnusson, C.~Berg, A.~Nielsen,
  H.~Skaug, M.~Machler, and B.~Bolker, ``{{glmmTMB}} balances speed and
  flexibility among packages for zero-inflated generalized linear mixed
  modeling,'' \emph{The R Journal}, 2017.

\bibitem{RbasePkg}
\BIBentryALTinterwordspacing
{R Core Team}, ``R: A language and environment for statistical computing.''
  [Online]. Available: \url{http://www.r-project.org/}
\BIBentrySTDinterwordspacing

\bibitem{modelbasedPkg}
D.~Makowski, D.~L{\"u}decke, M.~Ben-Shachar, and I.~Patil, ``{{modelbased}}:
  Estimation of model-based predictions, contrasts and means.''

\bibitem{paramatersArticle}
D.~L{\"u}decke, M.~Ben-Shachar, I.~Patil, and D.~Makowski, ``Extracting,
  computing and exploring the parameters of statistical models using {{R}},''
  \emph{J.\ of Open Source Software}, 2020.

\bibitem{parametersPkg}
\BIBentryALTinterwordspacing
D.~L{\"u}decke, D.~Makowski, M.~Ben-Shachar, I.~Patil, S.~H{\o}jsgaard, and
  B.~Wiernik, ``{{parameters}}: Processing of model parameters.'' [Online].
  Available: \url{https://CRAN.R-project.org/package=parameters}
\BIBentrySTDinterwordspacing

\bibitem{dplyrPkg}
\BIBentryALTinterwordspacing
H.~Wickham, R.~Fran{\c c}ois, L.~Henry, and K.~M{\"u}ller, ``{{dplyr}}: A
  grammar of data manipulation.'' [Online]. Available:
  \url{https://dplyr.tidyverse.org/}
\BIBentrySTDinterwordspacing

\bibitem{readrPkg}
\BIBentryALTinterwordspacing
H.~Wickham and J.~Hester, ``{{readr}}: Read rectangular text data.'' [Online].
  Available: \url{https://readr.tidyverse.org/}
\BIBentrySTDinterwordspacing

\bibitem{seeArticle}
D.~L{\"u}decke, I.~Patil, M.~Ben-Shachar, B.~Wiernik, P.~Waggoner, and
  D.~Makowski, ``{{see}}: An {{R}} package for visualizing statistical
  models,'' manuscript submitted for publication.

\bibitem{patchworkPkg}
\BIBentryALTinterwordspacing
T.~L. Pedersen, ``{{patchwork}}: The composer of plots.'' [Online]. Available:
  \url{https://CRAN.R-project.org/package=patchwork}
\BIBentrySTDinterwordspacing

\bibitem{sacha2015role}
D.~Sacha, H.~Senaratne, B.~C. Kwon, G.~Ellis, and D.~A. Keim, ``The role of
  uncertainty, awareness, and trust in visual analytics,'' \emph{IEEE
  transactions on visualization and computer graphics}, vol.~22, no.~1, pp.
  240--249, 2015.

\bibitem{etemadpour2014perception}
R.~Etemadpour, R.~Motta, J.~G. de~Souza~Paiva, R.~Minghim, M.~C.~F.
  De~Oliveira, and L.~Linsen, ``Perception-based evaluation of projection
  methods for multidimensional data visualization,'' \emph{IEEE Trans.\ on
  Visualization and Comp.\ Graphics}, 2014.

\end{thebibliography}
%

%

\begin{IEEEbiography}[{\includegraphics[width=1in,height=1.25in,clip,keepaspectratio]{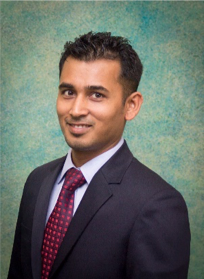}}]{Ghulam Jilani Quadri}
is a CIFellow postdoc at the University of North Carolina at Chapel Hill in the Department of Computer Science. He received his PhD degree from University of South Florida. Before joining the University of South Florida, Ghulam worked for Infosys Limited as System Engineer in Pune. Ghulam and team participated in IEEE VIS 2017 VAST Challenge and awarded Honorable mention. Ghulam received 2021 Computing Innovation Fellow award. His research interests are Information Visualization and HCI.
\end{IEEEbiography}

\begin{IEEEbiography}[{\includegraphics[width=1in,height=1.25in,clip,keepaspectratio]{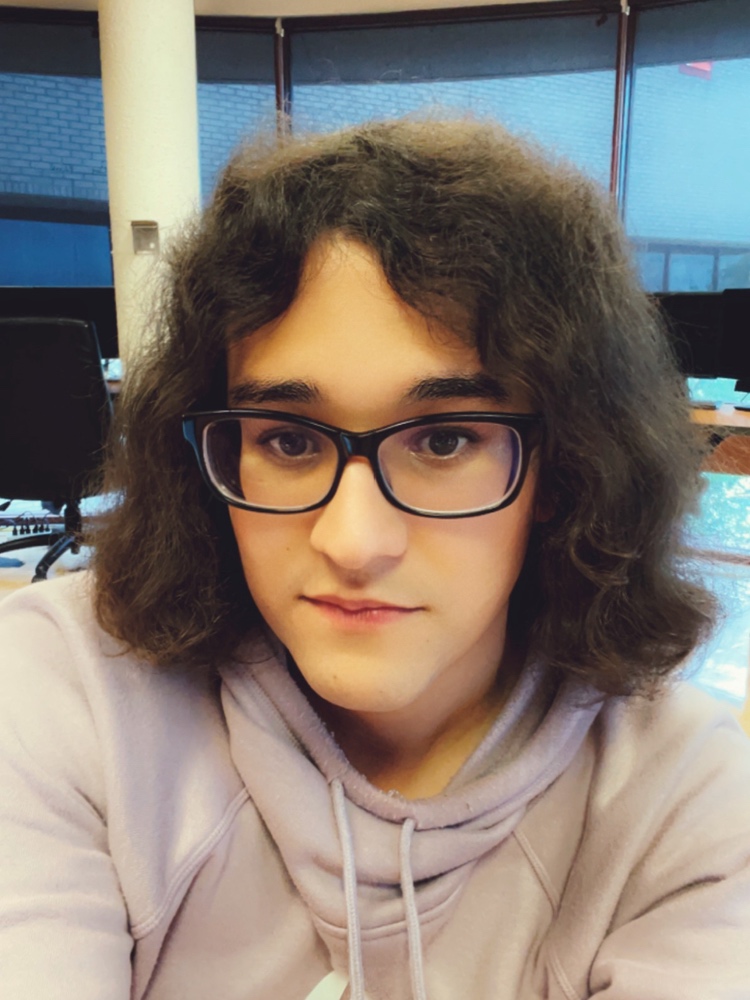}}]{Jennifer Adorno Nieves}
is a PhD Candidate of Computer Science and Engineering at the University of South Florida. Her dissertation work revolves around machine learning systems on public transportation for improving predicted arrival times as part of the Location Aware Information Systems Lab and in collaboration with the Center for Urban and Transportation Research. Additionally, she works with mobile technologies to develop cost effective approaches that can be used for diagnosis and remote monitoring of patients within the healthcare field. Generally, Jennifer's areas of interest include: Ubiquitous Sensing, Health and Accessibility. Over the summers, she serves as a mentor on the Research Experience for Undergraduates program hosted at her home institution.
\end{IEEEbiography}

\begin{IEEEbiography}[{\includegraphics[width=1in,height=1.25in,clip,keepaspectratio]{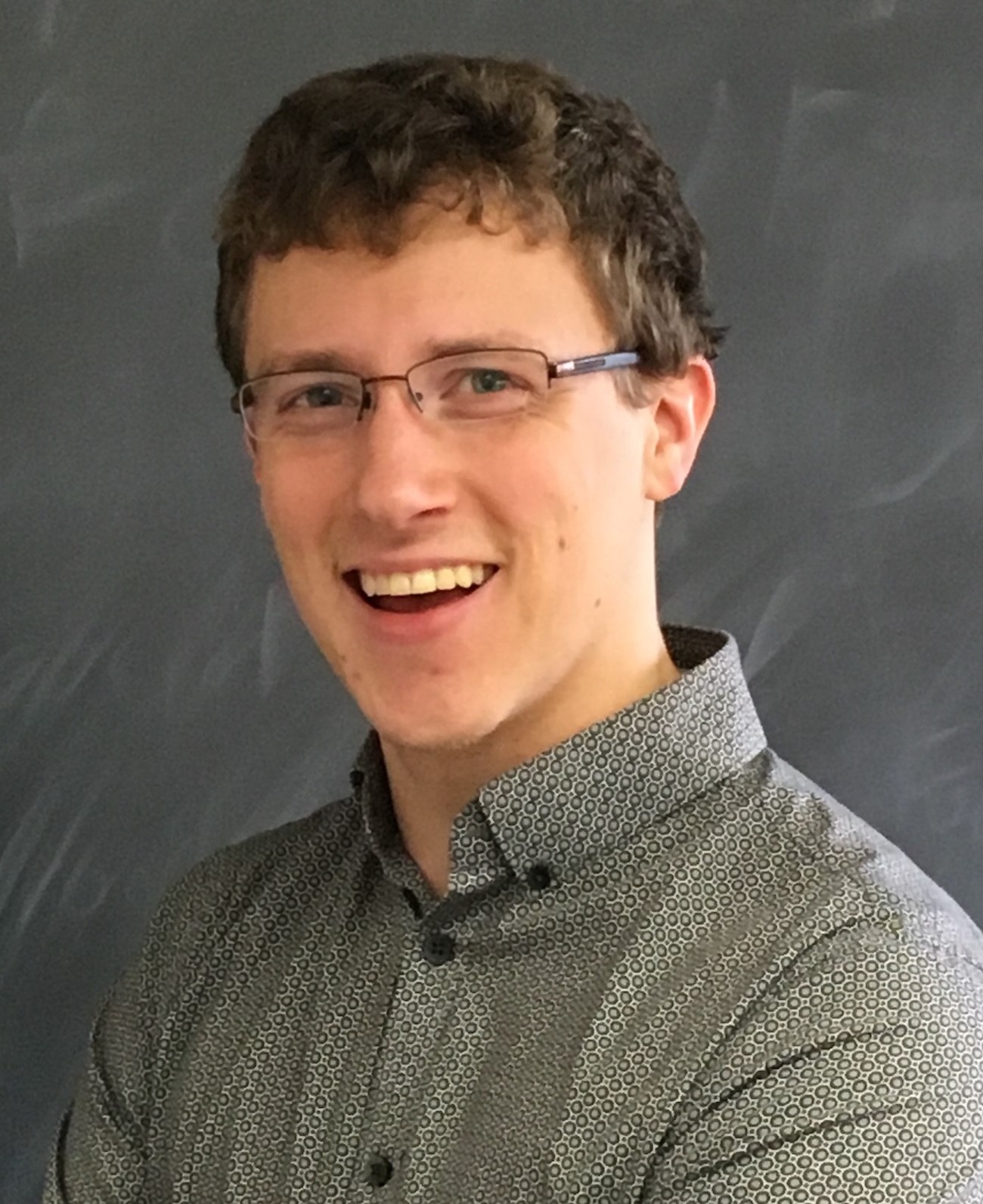}}]{Brenton M. Wiernik}
is an Affiliate Professor at the University of South Florida in the Department of Psychology. Dr. Wiernik is currently an independent researcher and Research Scientist at Meta, Demography and Survey Science. The current research was conducted while he was employed at the University of South Florida. He received his Ph.D.\ from the University of Minnesota. His recent research interests include psychometrics, latent variable modeling, and personality assessment. He is a developer of numerous R packages and Senior Editor for Organizational Behavior at Collabora: Psychology.
\end{IEEEbiography}

\begin{IEEEbiography}[{\includegraphics[width=1in,height=1.25in,clip,keepaspectratio]{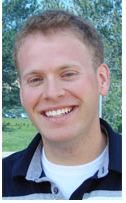}}]{Paul Rosen}
is an Associate Professor at the University of Utah. He received his Ph.D.\ from the Computer Science Department of Purdue University. His research interests include applying geometry- and topology-based approaches to problems in information visualization. Along with his collaborators, he has received awards for best paper at PacificVis 2016, IVAPP 2016, PacificVis 2014, and SIBGRAPI 2013.  Dr.\ Rosen received a National Science Foundation CAREER Award in 2019.
\end{IEEEbiography}




\end{document}